\documentstyle[12pt,equations]{article}
\input psfig

\def\MPL #1 #2 #3 {Mod.~Phys.~Lett.~{\bf#1},\  #2 (#3)}
\def\NPB #1 #2 #3 {Nucl.~Phys.~{\bf#1},\  #2 (#3)}
\def\PLB #1 #2 #3 {Phys.~Lett.~{\bf#1},\  #2 (#3)}
\def\PR #1 #2 #3 {Phys.~Rep.~{\bf#1},\ #2 (#3)}
\def\PRD #1 #2 #3 {Phys.~Rev.~{\bf#1},\  #2 (#3)}
\def\PRL #1 #2 #3 {Phys.~Rev.~Lett.~{\bf#1},\  #2 (#3)}
\def\RMP #1 #2 #3 {Rev.~Mod.~Phys.~{\bf#1},\  #2 (#3)}
\def\ZP #1 #2 #3 {Z.~Phys.~{\bf#1},\  #2 (#3)}
\def\IJMP #1 #2 #3 {Int.~J.~Mod.~Phys.~{\bf#1},\  #2 (#3)}

\def\siga{\sigma}
\def\ibid{{\it ibid.}}
\def\lh{{\rm LH}}
\def\half{{1\over 2}}
\def\chisq{\chi^2}
\def\cstar{\cos\theta^*}
\def\cstari{\cstar_1}
\def\cstarii{\cstar_2}
\def\mVV{M_{VV}}

\def\em{e^-}
\def\mup{\mu^+}
\def\mum{\mu^-}

\def\gam{\gamma}

\def\etal{{\it et al.}}

\def\anti{\overline}
\def\epem{e^+e^-}

\def\rts{\sqrt s}

\def\eg{{\it e.g.}}

\def\anti{\overline}
\def\wp{W^+}
\def\wm{W^-}
\def\wpm{W^{\pm}}
\def\mw{m_W}
\def\mz{m_Z}

\def\hsm{H}
\def\mhsm{m_{\hsm}}

\def\fbi{~{\rm fb}^{-1}}
\def\fb{~{\rm fb}}

\def\gev{~{\rm GeV}}
\def\tev{~{\rm TeV}}

\def\overlay#1#2{\ifmmode \setbox 0=\hbox {$#1$}\setbox 1=\hbox to\wd 0{\hss
$#2$\hss }\else \setbox 0=\hbox {#1}\setbox 1=\hbox to\wd 0{\hss #2\hss }\fi
#1\hskip -\wd 0\box 1}
\def\case#1/#2{{\textstyle{#1\over#2}}}
\def\9{\phantom 0}      
\renewcommand\linebreak{\unskip\break} 
\newcommand{\alt}{\mathrel{\raisebox{-.6ex}{$\stackrel{\textstyle<}{\sim}$}}}

\def\lsim{\alt}

\catcode`@=11
\newcount\@tempcntc
\def\@citex[#1]#2{\if@filesw\immediate\write\@auxout{\string\citation{#2}}\fi
  \@tempcnta\z@\@tempcntb\m@ne\def\@citea{}\@cite{\@for\@citeb:=#2\do
    {\@ifundefined
       {b@\@citeb}{\@citeo\@tempcntb\m@ne\@citea\def\@citea{,}{\bf ?}\@warning
       {Citation `\@citeb' on page \thepage \space undefined}}%
    {\setbox\z@\hbox{\global\@tempcntc0\csname b@\@citeb\endcsname\relax}%
     \ifnum\@tempcntc=\z@ \@citeo\@tempcntb\m@ne
       \@citea\def\@citea{,}\hbox{\csname b@\@citeb\endcsname}%
     \else
      \advance\@tempcntb\@ne
      \ifnum\@tempcntb=\@tempcntc
      \else\advance\@tempcntb\m@ne\@citeo
      \@tempcnta\@tempcntc\@tempcntb\@tempcntc\fi\fi}}\@citeo}{#1}}
\def\@citeo{\ifnum\@tempcnta>\@tempcntb\else\@citea\def\@citea{,}%
  \ifnum\@tempcnta=\@tempcntb\the\@tempcnta\else
   {\advance\@tempcnta\@ne\ifnum\@tempcnta=\@tempcntb \else \def\@citea{--}\fi
    \advance\@tempcnta\m@ne\the\@tempcnta\@citea\the\@tempcntb}\fi\fi}
\catcode`@=12

\renewenvironment{thebibliography}[1]
 {\begin{list}{\arabic{enumi}.}
    {\usecounter{enumi} \setlength{\parsep}{0pt}
     \setlength{\itemsep}{3pt} \settowidth{\labelwidth}{#1.}
     \sloppy
    }}{\end{list}}

\def\mm{\mu^+\mu^-}
\def\ee{e^+e^-}

\begin{document}
\thispagestyle{empty}

\newlength{\captsize} \let\captsize=\small 

%
\font\fortssbx=cmssbx10 scaled \magstep2
\hbox to \hsize{
%
%
$\vcenter{
\hbox{\fortssbx University of California - Davis}
\hbox{\fortssbx University of Wisconsin - Madison}
}$
\hfill
$\vcenter{
\hbox{\bf UCD-96-19} 
\hbox{\bf MADPH-96-949} 
\hbox{\bf IUHET-336}
\hbox{June 1996}
}$
}

\begin{center}
{\large\bf
Studying a Strongly Interacting Electroweak Sector
via  Longitudinal Gauge Boson Scattering
at a Muon Collider
}\\[.1in]
\small
V.~Barger$^a$, M.S.~Berger$^b$, J.F.~Gunion$^c$, T.~Han$^c$,
\\[.1in]
\small\it
$^a$Physics Department, University of Wisconsin, Madison, WI 53706,
USA\\
$^b$Physics Department, Indiana University, Bloomington, IN 47405,
USA\\
$^c$Physics Department, University of California,  Davis, CA 95616,  
USA\\
\end{center}

\vspace{.5in}

\begin{abstract}
We discuss the excellent prospects for a detailed study of a strongly
interacting electroweak sector at a muon collider with 
c.m. energy $\rts\sim 4\tev$.
For expected luminosity of $L=200-1000\fbi$ per year, $\mup\mum$ and
$\mup\mup$ (or $\mum\mum$) collisions can be used to study longitudinal
$\wp\wm$ and $\wp\wp$ (or $\wm\wm$) scattering with considerable precision.
In particular, detailed measurements of the distribution
in the $VV$ pair masses ($V=\wpm,Z$) will be possible. The shape and magnitude
of these distributions will provide a powerful tool for determining
the nature of strong gauge boson interactions.
Event rates will be large enough that projection techniques can be used
to directly isolate final states with different polarizations of the $V$'s
and verify that the strong interaction cross section excess is
mainly in the longitudinal-longitudinal mode.

\end{abstract}


\newpage
\setcounter{page}{1}
\pagenumbering{arabic}


\section{Introduction}
\indent

\begin{figure}[hb]
\let\normalsize=\captsize   
\begin{center}
\centerline{\psfig{file=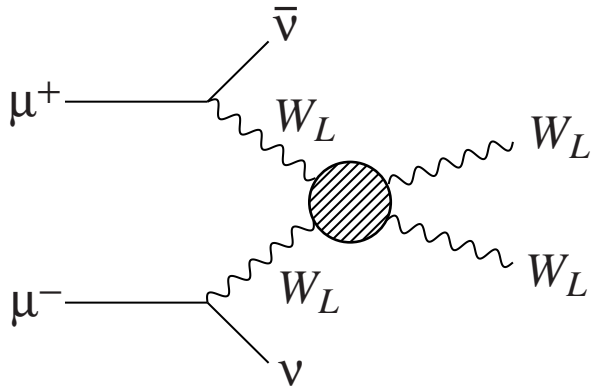,width=8cm}}
\begin{minipage}{12cm}       
\bigskip
\caption{{\baselineskip=0pt
Symbolic diagram for strong VV scattering.}}
\label{VVscatfig}
\end{minipage}
\end{center}
\end{figure}

Despite the extraordinary success of the Standard Model (SM) in
describing particle physics up to the highest energy available today,
the nature of electroweak symmetry-breaking (EWSB) remains undetermined.
In particular, it is conceivable that there is no light ($\lsim 700\gev$)
Higgs boson. General arguments \cite{strong} based on partial
wave unitarity then imply that the $\wpm,Z$ electroweak gauge
bosons  develop strong (non-perturbative) interactions 
by energy scales of order 1--2~TeV.
For a collider to probe such energy scales, the c.m. energy must be sufficient
that gauge-boson scattering (see Fig.~\ref{VVscatfig})
at subprocess energies at or above
1 TeV occurs with substantial frequency. The only colliders
under construction or being planned 
that potentially meet this requirement are the CERN LHC,
a next linear $\epem$ collider (NLC, with $\rts\lsim 1.5\tev$),
and a high energy muon collider (NMC).

The ability to extract signals and learn about
a strongly-interacting-electroweak sector (SEWS) at the LHC and
NLC has been the subject of many studies 
\cite{bbcghlry,LHCstudies,ATLASCMS,bchp,NLCstudies}.
The conclusion is that the LHC and NLC will
yield first evidence for a SEWS theory, but for many models
the evidence will be of rather marginal statistical significance.
SEWS models yielding large signals 
(such as the Standard Model with a $1\tev$ Higgs boson
or a model with a spin-1, isospin-1 resonance at $1\tev$) 
will be readily apparent or easily eliminated, but models
that yield only a small number of excess
events will be very difficult to distinguish from one another.
Measurement of the $VV$
mass spectrum (here we generically denote $\wpm,Z$ by $V$),
would reveal a wealth of information about SEWS models, but is
not feasible at the LHC or NLC.

Our focus in this paper is the ability of a muon collider to distinguish 
between and perform detailed studies of SEWS
models for longitudinal gauge-boson scattering. There may be
additional large and easily observed signals
in dynamical symmetry breaking models;
for example, in Technicolor models both Technicolor
hadrons \cite{ehlqkl} and techni-rho resonances \cite{bess,timb}
would be easily detected whenever the energy is adequate
for them to be produced at a reasonable rate. Only the SEWS signals are
addressed in the present work.

We shall demonstrate that a muon collider with center-of-mass energy, $\rts$,
of order 4 TeV would allow a comprehensive
study of the $\mVV$ distributions in all channels, assuming
(as should be the case) that $\mup\mum$ and $\mup\mup$ (or $\mum\mum$)
collisions are possible at the planned luminosity of
$L\sim 200-1000\fbi$ per year. A c.m. energy of $\rts=3-4\tev$
is the smallest that will allow such a detailed study.
Construction of a multi-TeV $\ee$ 
collider might also be a possibility \cite{dburke}, and 
would provide similar capabilities if an $\em\em$
facility is included, although bremsstrahlung-photon-initiated
backgrounds would be larger than at the muon collider.

In order to isolate the SEWS signals, it is necessary to determine
if there are events in the $\nu\anti\nu VV$ final
states at large $\mVV$ 
due to strong scattering of $V$'s with longitudinal ($L$) polarization
beyond those that will inevitably
be present due to standard electroweak processes,
including $VV$ scattering, that primarily
produce $V$'s with transverse ($T$) polarization.
There are two obvious ways of determining if such events are present.
\begin{itemize}
\item The first is to look an excess of events beyond what
is expected in the Standard Model when the Higgs boson
is light and there
is no strong scattering.  This involves reliably computing the
irreducible and reducible SM `backgrounds' and subtracting 
them from the observed rates.
\item The second is to employ projection techniques to separately
isolate the $V_LV_L$, $V_TV_T$ and $V_TV_L$ contributions.
\end{itemize}

Only the first procedure is practical at the LHC and NLC (with
$\rts\lsim 1.5\tev$) because of limited event rates.
The Standard Model with a light Higgs
boson of mass $\mhsm=100\gev$ is used to define the irreducible background,
and will be denoted by ``\lh''.\footnote{We note that
the specific choice of $\mhsm$ is not material so long as
it is well below the vector-boson pair threshold.}
This definition of the irreducible background
is appropriate since the growth in the $VV$
event rate in going from small $\mhsm$ to large $\mhsm$ 
(or because of some other SEWS model) is 
almost entirely due to an increase in the $V_LV_L$ rate,
the $V_TV_T$ and $V_TV_L$ rates being essentially independent 
of $\mhsm$. 
Thus, the SEWS signal is given by 
\begin{equation}
{d\Delta\sigma({\rm SEWS})\over d\mVV}\equiv {d\sigma({\rm SEWS})\over
d\mVV}-{d\sigma(\lh)\over d\mVV}\,,
\label{dsigdef}
\end{equation}
with $\Delta\sigma({\rm SEWS})$ being the integral thereof over a specified
range of $\mVV$.

At a $4\tev$ muon collider, the subtraction procedure
yields dramatic signals. Further, the projection
technique for isolating the longitudinal $V_L$ scattering rates
becomes very practical and exploration for new physics
beyond the SM becomes possible in all three polarization channels --- 
$TT$, $LL$ and $TL$ --- independently. As previously emphasized, 
in SEWS models the new strong interactions
affect only the $LL$ (and, in some cases, the $TL$) sector, and
not the $TT$ sector. However, theories predicting large anomalous
couplings could yield $TT$ rates that
also differ from SM expectations, and this difference could be uncovered
by the polarization analysis.

\section{Overview of Models}

\indent\indent
Numerous models for the strongly interacting gauge
sector have been considered. We focus on a selection
of those considered in Ref.~\cite{bbcghlry}:
\begin{itemize}

\item the Standard Model with a heavy Higgs boson of mass $\mhsm=1\tev$; 
\item a (``Scalar'') model in
which there is a spin-0, isospin-0 resonance with $M_S=1\tev$ but non-SM width
of $\Gamma_S=350\gev$;
\item a (``Vector'') model in which there is a spin-1, isospin-1
vector resonance with either 
$M_V=1\tev$ and $\Gamma_V=35\gev$ or $M_V=2\tev$ and $\Gamma_V=0.2\tev$,
but no spin-0 resonance. When necessary, we unitarize the model
using $K$-matrix techniques as detailed in the Appendix.
\item a model, denoted by LET-K or ``$\mhsm=\infty$'', in which 
the SM Higgs is taken to have infinite mass and the partial waves simply
follow the behavior predicted by the low-energy theorems, except that
the LET behavior is unitarized via the $K$-matrix techniques
described in the Appendix.
\end{itemize}
We note that the $\mhsm=1\tev$ Standard Model
is the simplest $VV$ scattering model for which the full kinematics 
and spin correlations among the final decay products are easily calculable.
Consequently, this model is extremely useful in bench
mark studies of the effectiveness of cuts and projection techniques.
As discussed in Ref.~\cite{bbcghlry}, 
the distributions of the final $V_L$'s, and their decay products, 
in other models
should follow closely those found for the $V_L$'s in the $\mhsm=1\tev$
Standard Model. Thus, in analyzing the other SEWS models,
we assume that cut efficiencies and
distributions are the same as for the $\mhsm=1\tev$ SM $V_L$'s.

Each distinct SEWS model yields a definite form for the
fundamental amplitude $A(s,t,u)$ defined by the weak isospin decomposition:
\def\CM{{\cal M}}
\begin{equation}
\CM(W^a_LW^b_L\rightarrow W^c_LW^d_L)\ =\ A(s,t,u)\delta^{ab}
\delta^{cd}\ +\ A(t,s,u)\delta^{ac}\delta^{bd}\ +
\ A(u,t,s)\delta^{ad}\delta^{bc}\ ,
\label{ispindecomp}
\end{equation}
where $a,b,c,d=1,2,3$, with $W_L^\pm=(1/\sqrt{2})(W_L^1 \mp iW_L^2)$ 
and $Z_L=W_L^3$.
All the physics of $V_LV_L$ scattering is contained in the
amplitude function $A(s,t,u)$. 
 
The physical amplitudes for
boson-boson scattering processes of interest are as follows:
\begin{eqnarray}
\CM(W^+_LW^-_L\rightarrow Z_LZ_L)\ &=&\ A(s,t,u) \\
\CM(W^+_LW^-_L\rightarrow W^+_LW^-_L)\ &=&\ A(s,t,u)\ +\ A(t,s,u) \\
\CM(W^\pm_LZ_L\rightarrow W^\pm_LZ_L)\ &=&\ A(t,s,u) \\
\CM(W^\pm_LW^\pm_L\rightarrow W^\pm_LW^\pm_L)\ &=&\ A(t,s,u)\ +
\ A(u,t,s)\ .
\label{channelforms}
\end{eqnarray}
These expressions for the amplitudes do not include the
symmetry factors for identical particles.
The isospin amplitudes $T_I$, for isospin $I$, are given by
\begin{eqnarray}
T_0\ & =&\ 3\,A(s,t,u)\ +\ A(t,s,u)\ +\ A(u,t,s)\,,  \\
T_1\ & =&\ A(t,s,u)\ -\ A(u,t,s)  \,,\\
T_2\ & =&\ A(t,s,u)\ +\ A(u,t,s) \,.
\label{isospinamps}
\end{eqnarray}
In terms of the $T_I$, the relevant physical scattering
amplitudes can be written as
\begin{eqnarray}
\CM(W^+_LW^-_L\rightarrow Z_LZ_L)\ &=&\  {1\over 3}[ T_0\ -\ T_2] 
\label{wwisospini} \\
\CM(W^+_LW^-_L\rightarrow W^+_LW^-_L)\ &=&\  {1\over6}[2T_0\ +\ 3T_1\ +
\ T_2] \label{wwisospinii} \\
\CM(W^\pm_LZ_L\rightarrow W^\pm_LZ_L)\ &=&\  {1\over2}[T_1\ +\ T_2]  
\label{wwisospiniii} \\
\CM(W^\pm_LW^\pm_L\rightarrow W^\pm_LW^\pm_L)\ &=&\  T_2\ .
\label{wwisospiniv}
\end{eqnarray}
Again, these amplitudes do not include the identical particle symmetry factors.

Measurements of the processes
\begin{eqnarray}
\wp\wm&\to&\wp\wm,ZZ\;, \nonumber \\
\wpm Z &\to & \wpm  Z\;, \label{processlist} \\
\wpm\wpm &\to &\wpm\wpm \;, \nonumber
\end{eqnarray}
as a function of $s$, $t$, and $u$ 
provide as much information on the $T_I$ and the function
$A(s,t,u)$ as can be accessed experimentally.
A full reconstruction of the $T_I$, including phases, is not 
possible since the cross sections depend upon the amplitudes of
Eqs.~(\ref{wwisospini}-\ref{wwisospiniv}) squared. 
If it is necessary to integrate
over the $s,t,u$ variables in order to obtain statistically significant
measurements, then much information is lost. 

As shown below, a $4\tev$ muon collider can provide at least a
reasonably good determination of the $\mVV$ invariant mass
spectrum for each of the above reactions.
In contrast, for most models, the LHC
or a $\rts\lsim 1.5\tev$ NLC can at best allow determination
of integrals over broad ranges of $\mVV$.

\section{Probing SEWS Models at the LHC and NLC}

\indent\indent
If the electroweak sector is strongly interacting,
partial exploration of the underlying SEWS model in the 
three weak-isospin channels ($I=0,1,2$) will be possible at the LHC.
The signal and background for gold-plated (purely leptonic) events is shown
in Table~\ref{tableiii} for the LHC operating at $14\tev$ with $L=100\fbi$,
for several of the above models. 
These event rates were computed after imposing a series 
of crucial cuts required to suppress background
and after integrating over a broad range
of $\mVV$ values (or effectively so for final states in
which $\mVV$ cannot be directly reconstructed).
Discrimination among models
is achieved by comparing the gold-plated event rates in the different
channels. For example, if the signal rate
is largest in the $\wpm\wpm$ like-sign channels, then non-resonant
models such as the LET-K would be preferred. Similarly, a large $\wpm Z$
signal relative to the $ZZ$ and $\wpm\wpm$ channels would favor
a Vector resonance model. However, only in the case of the $M_V=1\tev$
Vector model would there be any chance of actually observing
details regarding the structure of the $\mVV$ spectrum at the LHC.
A $M_V=2\tev$ Vector model would be virtually indistinguishable from
the LET-K model.

\begin{table}[h]
\centering
\caption[]{\label{tableiii}\small
Total numbers of $V_LV_L \to 4$-lepton
signal $S$ and background $B$ events calculated for the
LHC \protect\cite{bbcghlry}, assuming $L=100\fbi$.
The Vector model is that with $M_V=1\tev$ and $\Gamma_V=35\gev$.
The Scalar model has $M_S=1\tev$, $\Gamma_S=350\gev$.}
\bigskip
\begin{tabular}{|c|c|c|c|c|}
\hline
& Bkgd & Scalar & Vector & LET-K \\ 
\hline \hline
$ZZ (4\ell)$& 1 & 5 & 1.5 & 1.5\\
$(2\ell 2\nu)$& 2 & 17 & 5 & 4.5\\ \hline
$W^+W^-$& 12 & 18 & 6 & 5\\ \hline
$W^+Z$& 22 & 2 & 70 & 3\\ \hline
$W^{\pm }W^{\pm }$& 4 & 7 & 12 & 13\\ \hline
\end{tabular}
\end{table}

The channels and models in Table~\ref{tableiii}
have also been studied for a $1.5\tev$ NLC \cite{bchp}.
As illustrated in Table~\ref{tablenlc}, event rates in the (usable) four jet
final states are more promising than in the purely leptonic final states
at the LHC. The rates are at a level that SEWS models with large
signals, such as the $\mhsm=1\tev$ SM or $m_V=1\tev$ Vector model,
would be apparent, but detecting the signals for models
without dramatic resonances, such as the LET-K model, would be difficult.
It would also not be possible to distinguish the $\mhsm=1\tev$
model from a Scalar model with narrower resonance width.
Detailed study of the strong interactions through the $\mVV$
distributions would not be possible.

\begin{table}
\def\arraystretch{.8} 
\centering
\smallskip
\caption[]{\label{tablenlc} \small
Total numbers of $W^+W^-, ZZ \rightarrow  4$-jet
signal $S$ and background $B$ events calculated for  a 1.5~TeV
NLC with  integrated luminosity 200~fb$^{-1}$,
using $100\%$ polarized $e^-_L$ beams \cite{bchp}.
Events are summed
over the mass range $0.5 < M_{WW} < 1.5$~TeV except for the $W^+W^-$ channel
with  a narrow vector resonance in which $0.9 < M_{WW} < 1.1$~TeV. The
statistical significance $S/\sqrt B$ is also given.
For comparison, results for $e^-e^- \rightarrow \nu \nu W^-W^-$
are also presented, for the same energy and luminosity and the $W^+W^-$
cuts. The hadronic branching fractions of $WW$ decays and the $W^\pm/Z$
identification/misidentification are included.
}
\bigskip
\begin{tabular}{|l|c|c|c|c|}
\hline
channels & SM  & Scalar & Vector   & LET-K  \\
& $m_H=1$ TeV & $M_S=1$ TeV & $M_V=1$ TeV &\\
\hline\hline
$S(e^+ e^- \to \bar \nu \nu W^+ W^-)$
& 330   & 320   & 92  & 62  \\
$B$(backgrounds)
& 280    & 280   & 7.1  & 280  \\
$S/\sqrt B$ & 20 & 20 & 35 & 3.7 \\
\hline
$S(e^+ e^- \to \bar\nu \nu ZZ)$
&  240  & 260  & 72  & 90   \\
$B$(backgrounds)
& 110    & 110   & 110  & 110  \\
$S/\sqrt B$ & 23 & 25& 6.8& 8.5\\
\hline
\hline
$S(e^- e^-_L \to \nu \nu W^- W^-)$  & 54 & 70 & 72 & 84 \\
$B$(background) & 400 & 400 & 400 & 400\\
$S/\sqrt B$ & 2.7 & 3.5 & 3.6 & 4.2 \\
\hline
$S(e^-_L e^-_L \to \nu \nu W^- W^-)$  & 110 & 140 & 140 & 170 \\
$B$(background) & 710 & 710 & 710 & 710\\
$S/\sqrt B$ & 4.0 & 5.2 & 5.4 & 6.3 \\ \hline
\end{tabular}
\end{table}

\section{Rates and Motivations for Higher Energy}

\begin{figure}
\let\normalsize=\captsize   
\begin{center}
\centerline{\psfig{file=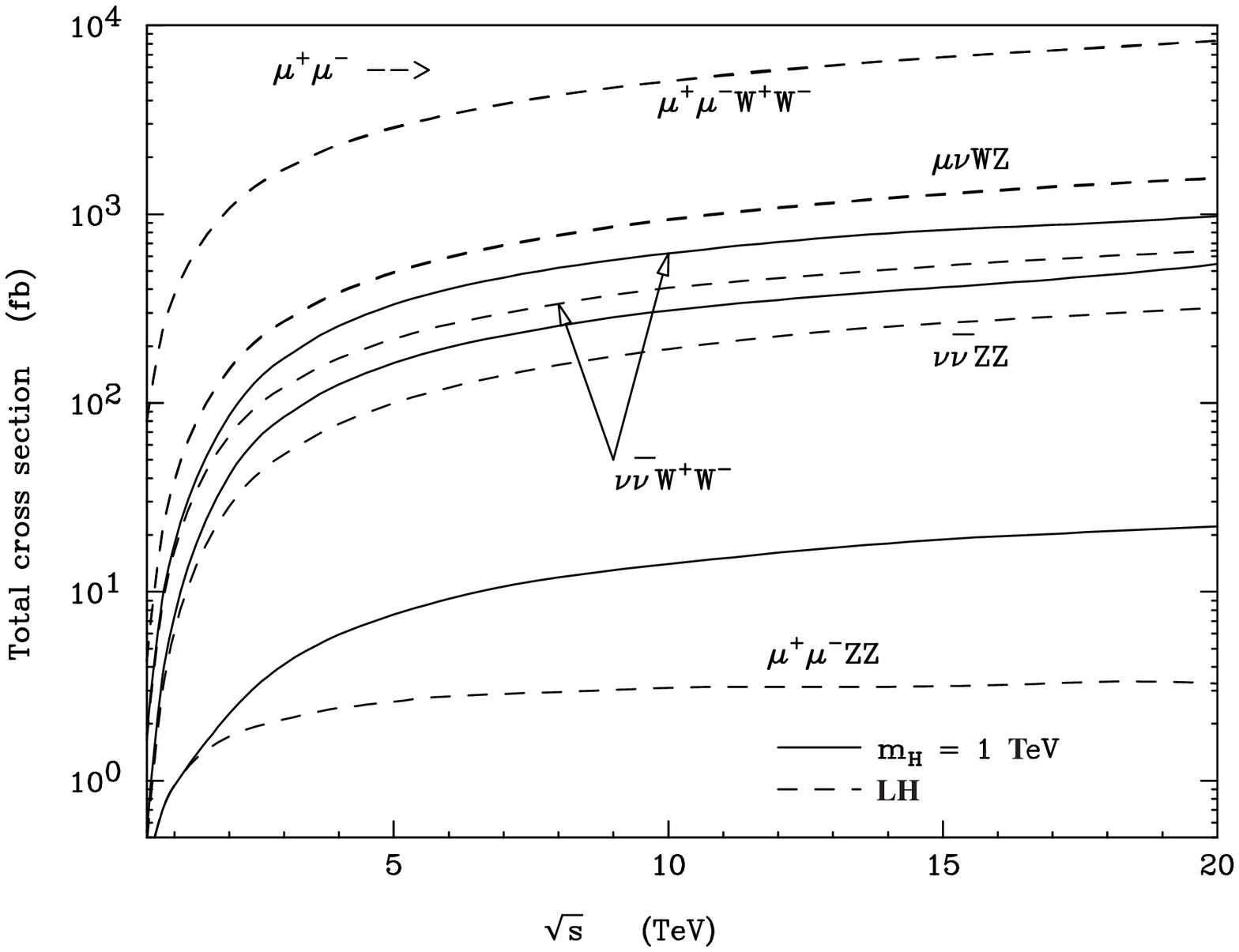,width=12cm}}
\begin{minipage}{12cm}       
\bigskip
\caption{{\baselineskip=0pt
Signal and background cross sections as a function of
$\protect\rts$ for strong $\wp\wm\to\wp\wm$
and $\wp\wm\to ZZ$ scattering as computed in the SM
for $\mhsm=1\tev$ at a $\mup\mum$ collider.}}
\label{sewssm}
\end{minipage}
\end{center}
\end{figure}

\indent\indent
For a first estimate of the strong electroweak scattering effects
we take the Standard Model with a heavy Higgs as a prototype of
the strong scattering sector. 
For a 1 TeV SM Higgs boson, the SEWS signal is accordingly defined as
\begin{equation}
\Delta \sigma =\sigma(\mhsm=1\tev)-\sigma(\lh)\;.
\label{signaldef}
\end{equation}
Results in the $\wp\wm$ and $ZZ$ channels
for $\Delta\sigma$ (with no cuts of any kind)
are shown in Table~\ref{tableii}
for $\rts=1.5\tev$ (as often discussed
for an $\ee$ collider) and $4\tev$. The strong 
scattering signal is relatively small at energies of order $1\tev$, but 
grows substantially as multi-TeV energies are reached.
This is illustrated in Fig.~\ref{sewssm},
where $\sigma(\mhsm=1\tev)$ and $\sigma(\lh)$ are plotted
separately. The signal $\Delta\sigma$ declines rapidly in magnitude
for decreasing energy below $\rts=4\tev$. The associated signal ($S$) and
irreducible background event rates are given by $S=L\Delta\sigma$
and $L\sigma(\lh)$, where $L$ is the integrated luminosity.
Table~\ref{tableii} shows that a very respectable signal rate is achieved 
at $4\tev$ and even before cuts the signal to 
irreducible background ratio is quite reasonable; both are
much larger than at $\rts=1.5\tev$.
SEWS physics benefits from increasing energy in four ways.
\begin{itemize}
\item
The luminosity for $V_LV_L$ collisions is bremsstrahlung-initiated
and grows at fixed subprocess $\hat s=\mVV^2$
as $1/\tau$ where $\tau=\mVV^2/s$.
\item
The SEWS amplitude function $A(\hat s,\hat t,\hat u)$ typically increases
as higher subprocess $\hat s=\mVV^2$ values become accessible;
\eg\ in the LET-K model $A(\hat s,\hat t,\hat u)\propto \hat s/v^2$,
where $v$ is the standard electroweak symmetry breaking parameter.
This more than compensates for the slightly faster growth with $s$
of the $V_TV_T$ luminosity function (responsible
for $VV$ fusion backgrounds) $\propto{1\over\tau}\ln^2{s\over m_V^2}$
as compared to the $V_LV_L$ luminosity.
\item
The background subprocess
amplitudes typically have point-like $1/\hat s$ behavior
and, further, some backgrounds
are not proportional to the growing $VV$ luminosities.
In particular, many of the diagrams contributing to the amplitude
for the irreducible light Higgs $V_TV_T+V_TV_L$ SM background 
do not have $VV$ fusion topology.
\item
Finally, the luminosities at higher machine energies
are normally designed to be larger to compensate for the $1/\hat s$ decline
of the point-like subprocess cross sections for other types of new physics.
\end{itemize}
It appears that $\rts=4\tev$ 
is roughly the critical energy at which SEWS physics can
first be studied in detail. This is especially true given that it
will be desirable to impose strong cuts in order to maximize 
signal over background.
Thus, the high energy reach of a muon collider could prove
to be critically important.

\begin{table}[htb]
\centering
\caption[]{\label{tableii}\small
Strong electroweak scattering signals in $\wp\wm\to\wp\wm$
and $\wp\wm\to ZZ$ at future lepton colliders.
Signal cross sections and the signal to irreducible \lh\ background ratio
are given for $L=200\fbi$ at $\rts=1.5\tev$
and $L=1000\fbi$ at $\rts=4\tev$, with no cuts.}
\medskip
\begin{tabular}{|c|c|c|c|c|}
\hline
$\sqrt s$& $\Delta\sigma(W^+W^-)$& $[S/\lh](\wp\wm)$&
$\Delta\sigma(ZZ)$ &$ [S/\lh](ZZ)$\\ \hline \hline
1.5 TeV& 8 fb& ${1600\over 8000}$ & 6 fb & ${1200\over 3600}$ \\ \hline
4 TeV& 80 fb & ${80000\over 170000}$ & 50 fb & ${50000\over 80000}$
\\ \hline
\end{tabular}
\end{table}

\begin{table}[htb]
\centering
\caption[]{\label{polarizationtable}\small
Standard Model cross sections (in $\fb$) 
at $\protect\rts=4\tev$ in the $\wp\wp$ final state
$TT$, $TL$ and $LL$ modes for a light Higgs compared to $\mhsm=1\tev$.
Results are given both without any cuts and after imposing cuts I and II
on the $\wp$'s as delineated in Section 4.}
\medskip
\begin{tabular}{|c|cccc|cccc|}
\hline
\ & \multicolumn{4}{c|}{\lh} & \multicolumn{4}{c|}{$\mhsm=1\tev$}
\\
\ & $TT$ & $TL$ & $LL$ & Sum & $TT$ & $TL$ & $LL$ & Sum \\ \hline\hline
No Cuts & 48.9 & 26.9 & 6.75 & 82.6 & 49.5 & 26.5 & 12.1 & 88.1 \\ \hline
With Cuts & 1.34 & 0.19 & 0.03 & 1.56 & 1.36 & 0.15 & 1.51 & 3.02 \\ \hline
\end{tabular}
\end{table}

The importance of high energy and signal selection
cuts is particularly apparent in 
the $\wp\wp$ channel.  We choose this
channel to illustrate the polarization structure of the 
$VV$ final state.  In Table~\ref{polarizationtable}
we compare the polarization decomposition of the 
Standard Model $\wp\wp$ cross sections for the \lh\ case with the
$\mhsm=1\tev$ SEWS model. (We define the vector boson polarizations
in the $VV$ rest frame.) 
With no cuts, substantial $TL$ and not insignificant
$LL$ cross section components are present for the \lh\ model,
in addition to the dominant $TT$ cross section.
The increase ($\Delta\sigma$) in
cross section in going to $\mhsm=1\tev$ is a small
percentage of the total and is seen to be almost entirely
in the $LL$ final state (the $TL$ contribution actually decreases).
By imposing cuts on the $\wp$'s (specifically
cuts I and II as delineated in Section 4),
the $TL$ and, especially, $LL$ cross sections are reduced to
negligible size compared to $TT$ for the \lh\ model.
Further, with the cuts imposed, $\Delta\sigma$ 
is nearly as large as $\sigma(\lh)$. Thus,
cuts are important both in reducing 
the background relative to the signal and also in isolating
the $LL$ component of the strongly interacting gauge boson cross section.
Whether cuts are imposed or not, it is clear from 
Table~\ref{polarizationtable} that 
the $TT$ contribution to $\Delta\sigma$ is negligible.
Finally, the magnitude of $\Delta\sigma$ in the $\wp\wp$ channel
(even before cuts) becomes sufficient at $\rts=4\tev$ to allow
quantitative study; for $\rts\lsim 1.5\tev$ $\Delta\sigma$ is 
so small that the $\wp\wp$ channel can at best provide only
a hint of strong interactions among the gauge bosons.

\section{Muon Collider Results using the Subtraction Procedure}

\indent\indent
For a $\mm$ collider operating at $4\tev$ the event rates
and statistical significances for most channels
markedly improve, the exception being the $\wpm Z\to\wpm Z$ channel.
Table~\ref{tableiv} summarizes 
our results for various SEWS models in the $\wp\wm\to \wp\wm$,
$\wp\wm\to ZZ$ and $\wp\wp\to\wp\wp$ channels\footnote{We focus
on the $\wp\wp\to\wp\wp$ like-sign channel, but the
same results apply to the $\wm\wm\to\wm\wm$ channel.
High luminosity $\mup\mup$ collisions may be somewhat
easier to achieve.}
for the signal $S$ and the full irreducible plus reducible 
background $B$ event numbers obtained by summing over diboson invariant mass 
bins as specified in the caption. Also given is
the statistical significance, $S/\sqrt B$, of the signals.
The signal rate $S$ is computed
by subtracting the background rate $B$ from the
total event rate (signal+background) for a given SEWS model,
see Eq.~(\ref{signaldef}).
The results presented in Table~\ref{tableiv} are those obtained after
imposing the following cuts:
\begin{description}
\item{I:} Basic Cuts: $p_T(V)\geq 150\gev$; $|\cos\theta_V|\leq 0.8$;
$\mVV\geq 500\gev$; 
\item{II:} $20\gev\leq p_T(ZZ)\leq 300\gev$; $30\gev\leq p_T(WW)\leq 300\gev$;
and veto of any $\mu^\pm$ with $\theta_\mu\geq 12^\circ$ and $E_\mu\geq
50\gev$ (assuming the beam hole has a $12^\circ$ opening);
\item{III:} separation of $W$'s from $Z$'s in the 4-jet final state
using the mass cuts in Eq.~(\ref{masscuts}) below.
\end{description}
The $p_T(VV)$ cuts and muon veto in II are designed to suppress
reactions such as $\mup\mum\to\mup\mum \wp\wm$
deriving from subprocesses involving initial
state photons emitted from the incoming $\mup,\mum$.
The veto eliminates a large fraction of such events.
Further, the $V_LV_L$ signal tends to have $p_T(VV)\sim \mw$, whereas
$p_T(VV)$ for the $\mup\mum\wp\wm$
final state is quite small, especially after the veto cut.
In order to separate $W$'s from $Z$'s in the final state
we define, following Ref.~\cite{bchp}, a $W$, $Z$ as two jets having
invariant mass in the ranges
\begin{equation}
M^W_{jj}\in [0.85\mw,\half(\mw+\mz)]\,,\quad M_{jj}^Z\in
[\half(\mw+\mz),1.15\mz]\,,
\label{masscuts}
\end{equation}
respectively.
For a detector with resolution $\Delta E_j/E_j=0.50/\sqrt {E_j}\oplus 0.02$,
the true $WW$, $WZ$ and $ZZ$ final states will be interpreted statistically
as follows:
\begin{eqnarray}
WW&\Rightarrow& 78\%~WW,~18\%~WZ,~~1\%~ZZ,~3\%~{\rm reject}\,,\nonumber\\
WZ&\Rightarrow& 11\%~WW,~77\%~WZ,~~9\%~ZZ,~3\%~{\rm reject}\,,\nonumber\\
ZZ&\Rightarrow& ~7\%~WW,~22\%~WZ,~72\%~ZZ,~4\%~{\rm reject}\,.\nonumber
\end{eqnarray}
Misidentification of a $WW$ final state as $ZZ$ is especially unlikely.

\begin{table*}
\centering
\caption[]{\label{tableiv}\small
Total numbers of $W^+W^-, ZZ$ and $W^+W^+ \rightarrow4$-jet
signal ($S$) and background ($B$) events calculated for  a 4~TeV
$\protect \mm$ collider with  integrated luminosity 200~fb$^{-1}$
(1000~fb$^{-1}$ in the parentheses), for cuts of $\mVV\geq 500\gev$,
$p_T(V)\geq 150\gev$, $|\cos\theta_V|\leq 0.8$ and $p_T(WW)\geq 30\gev$,
$p_T(ZZ)\geq 20\gev$. (For the case of a 2 TeV
vector state, events for the $W^+W^-$ channel are summed around
the mass peak over the range $1.7 < \mVV < 2.3$~TeV.)
Events containing a $\mup$ or $\mum$ with
$\theta_\mu\geq 12^\circ$ and $E_\mu\geq 50\gev$ are vetoed.
The signal rate $S$ is that obtained by computing the total rate
(including all backgrounds) for a given SEWS model and then
subtracting the background rate; see Eq.~(\ref{signaldef}).
The statistical significance $S/\sqrt B$ is given for the signal
from each model.
The hadronic branching fractions of $VV$ decays and the $W^\pm/Z$
identification/misidentification are included.}
\bigskip
\begin{tabular}{|l|c|c|c|} \hline
 & Scalar  & Vector & LET-K  \\
\noalign{\vskip-1ex}
& $m_H=1$ TeV  & $M_V=2$ TeV & $\mhsm=\infty$ \\
channels & $\Gamma_H=0.5$ TeV  & $\Gamma_V=0.2$ TeV & Unitarized\\ 
\hline\hline
$\mu^+ \mu^- \to \bar \nu \nu W^+ W^-$ &  &  & \\
$S$(signal) & 2400 (12000)  & 180 (890)  & 370 (1800)  \\
$B$(backgrounds)
& 1200 (6100)   & 25 (120)  & 1200 (6100)  \\
$S/\sqrt B$ & 68 (152) & 36 (81) & 11 (24) \\ \hline
$\mu^+ \mu^- \to \bar\nu \nu ZZ$& & &  \\
$S$(signal) &  1030 (5100)  & 360 (1800)  & 400  (2000)  \\
$B$(backgrounds)
& 160 (800)    & 160 (800)  & 160 (800) \\
$S/\sqrt B$ & 81 (180) & 28 (64) & 32 (71) \\ \hline
$\mu^+ \mu^+ \to \bar\nu \bar\nu W^+ W^+$ & &  &  \\
$S$(signal) & 240 (1200) & 530 (2500) & 640 (3200)   \\
$B$(backgrounds)
& 1300 (6400)  & 1300 (6400)  & 1300 (6400) \\
$S/\sqrt B$ & 7 (15) & 15 (33) & 18 (40) \\ \hline
\end{tabular}
\end{table*}

Before turning to an examination of the $\mVV$ distributions,
we summarize the implications of the integrated
event rates appearing in Table~\ref{tableiv}.
Most importantly, the statistical significance of the SEWS signal
is high for all channels, regardless of model.
Even the  $\mhsm=\infty$, $\wp\wm$ and $\mhsm=1\tev$, $\wp\wp$ 
signals are clearly visible with only $L=200\fbi$, 
becoming thoroughly robust for $L=1000\fbi$. (Note that
in this case the ratio $S/B$ can be enhanced by making a higher
mass cut (\eg\ $M_{VV} > 0.7$ TeV), but the significance $S/\sqrt B$
is not improved.)
Second, models of distinctly different types are easily distinguished
from one another.
A broad Higgs-like scalar will enhance both $W^+ W^-$
and $ZZ$ channels with $\sigma(W^+ W^-) > \sigma(ZZ)$; a $\rho$-like vector
resonance will manifest itself through $W^+W^-$ but not $ZZ$; the 
unitarized $\mhsm=\infty$ (LET-K)
amplitude will enhance $ZZ$ more than $W^+ W^-$.

\begin{figure}[tbp]
\let\normalsize=\captsize   
\centering
\centerline{\psfig{file=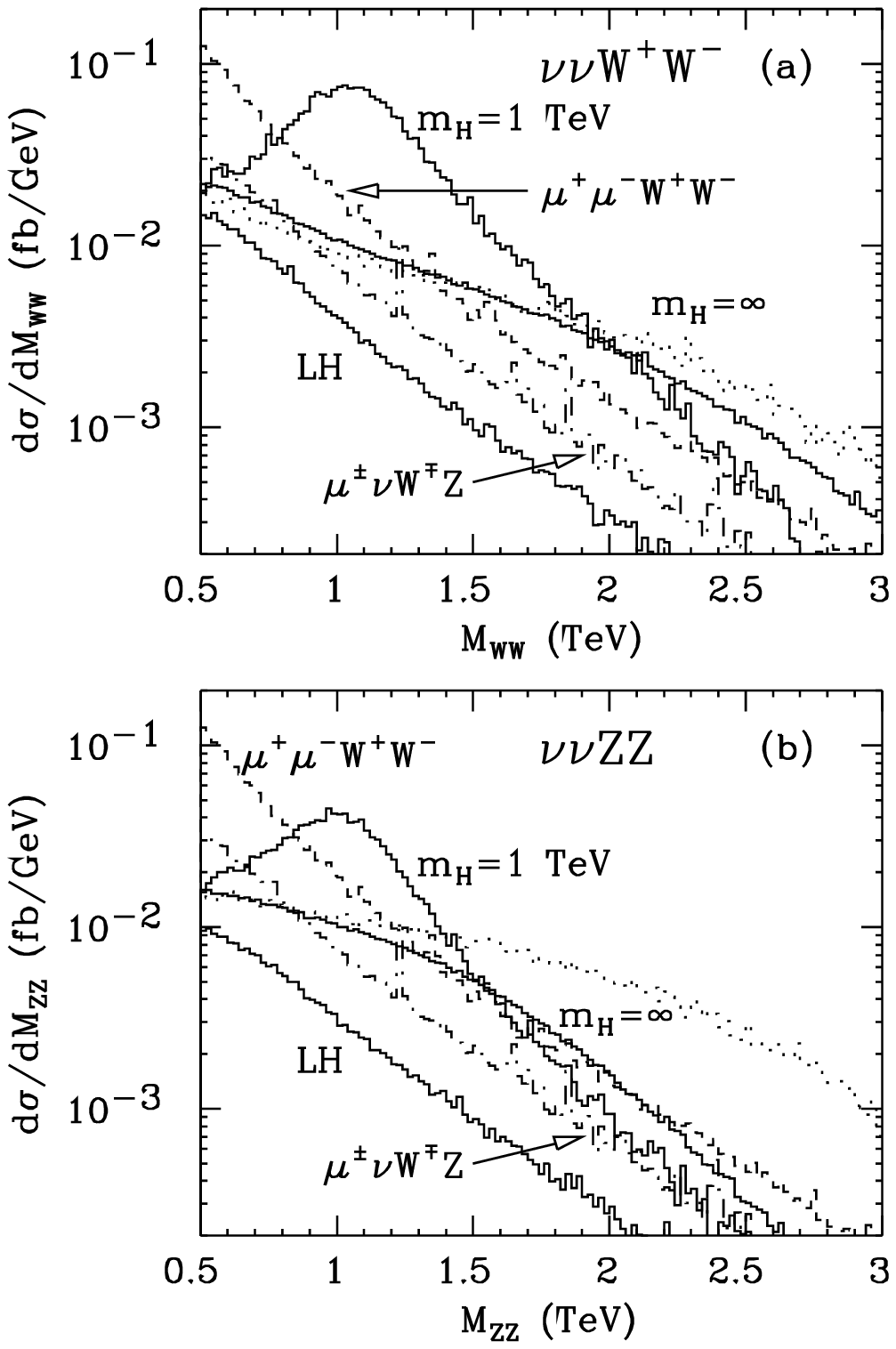,width=10cm}}
\begin{minipage}{12cm}       
\smallskip
\caption{{\baselineskip=0pt
Cross section as a function of $\mVV$
for SEWS models and backgrounds at $\protect\rts=4\tev$ in the 
(a) $\wp\wm$ and (b) $ZZ$ final states
after imposing only the basic cuts I. The irreducible
background is given by the 
strictly electroweak \lh\ limit of the Standard Model. 
The most important
reducible backgrounds, $\mup\mum\to\mup\mum\wp\wm$
and $\mup\mum\to\mu^\pm\nu W^\mp Z$, are also displayed.
The $\mVV$ distributions (without subtracting
the \lh\ background) for two sample SEWS models are shown:
(i) the SM Higgs with $\mhsm=1\tev$;
and (ii) the SM with $\mhsm=\infty$ unitarized via $K$-matrix techniques
(LET-K model). Also shown by the dotted histogram 
is the $\mhsm=\infty$ SM result before $K$-matrix unitarization.}}
\label{mvvbasic}
\end{minipage}
\end{figure}

\begin{figure}[tbp]
\let\normalsize=\captsize   
\centering
\centerline{\psfig{file=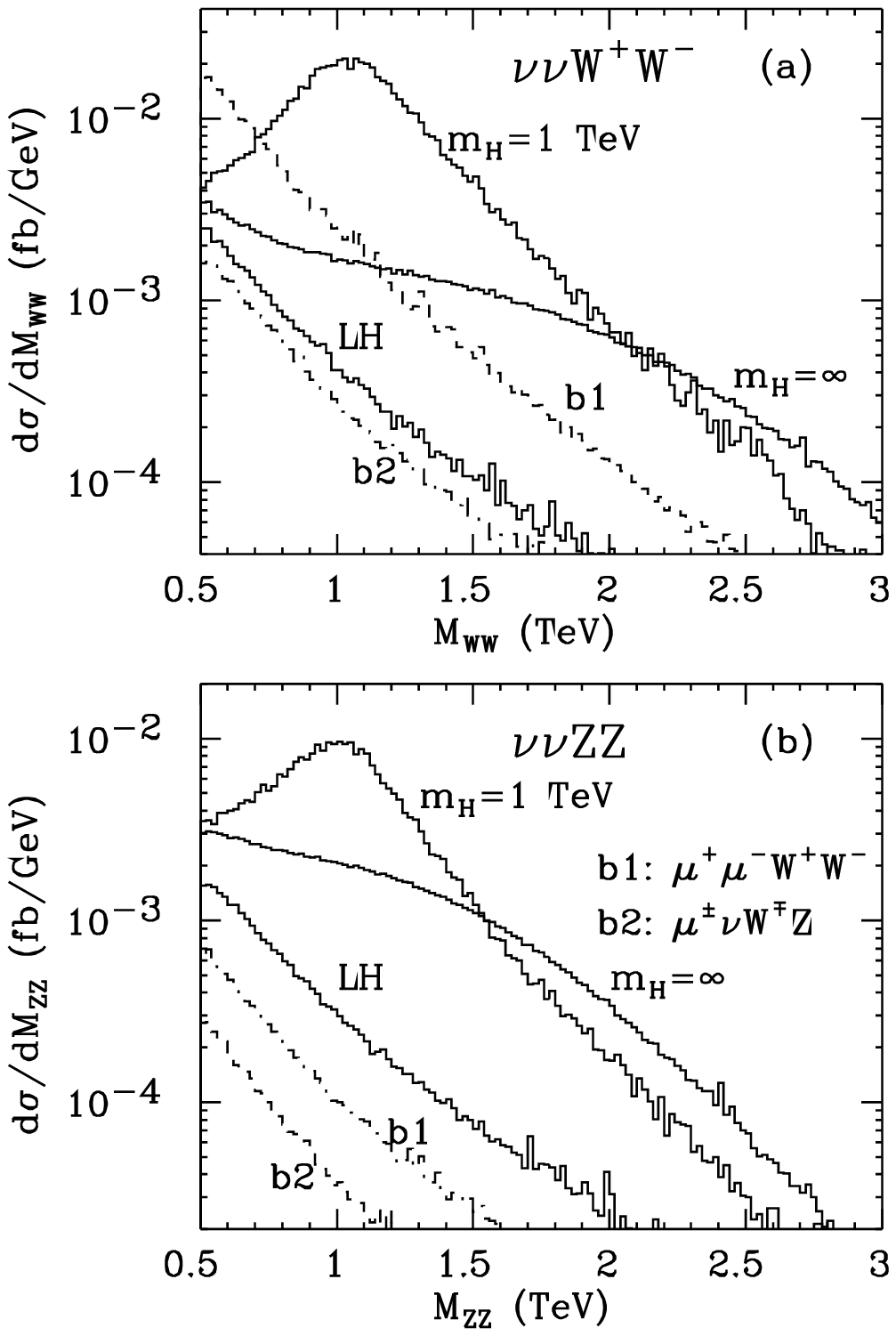,width=10cm}}
\begin{minipage}{12cm}       
\smallskip
\caption{{\baselineskip=0pt
Differential cross sections at $\protect\rts=4\tev$ versus $\mVV$
for SEWS models and backgrounds in the 
(a) $\wp\wm$ and (b) $ZZ$ final states
after imposing all cuts I--III. See caption for Fig.~\ref{mvvbasic}
for a description of the models.}}
\label{mvvcuts}
\end{minipage}
\end{figure}

The importance of the $p_T(VV)$ cut, the veto of energetic muons outside
the beam hole and the mass cuts is illustrated by
comparing Fig.~\ref{mvvbasic} to Fig.~\ref{mvvcuts}.
In Fig.~\ref{mvvbasic},
we plot the $\mhsm=1\tev$ and unitarized $\mhsm=\infty$
signals, the irreducible \lh\ background
and the most important reducible backgrounds after imposing
only the basic cuts I. Without the additional cuts II and III the reducible
backgrounds are much more important than the irreducible $\nu\anti\nu \wp\wm$
and $\nu\anti\nu ZZ$
backgrounds.  Also evident from this figure is the importance
of unitarizing the $\mhsm=\infty$ partial wave behavior predicted
in the SM.  Without unitarization (the dotted histograms) there
is an extensive (unphysical) tail at high $\mVV$.
These histograms should be compared
to the corresponding histograms in Fig.~\ref{mvvcuts}
obtained after imposing all the cuts, I-III.  The dramatic
reduction in the reducible backgrounds relative to the SEWS
signals and the irreducible \lh\ background
in both the $\wp\wm$ and $ZZ$ channels offsets some 
sacrifice in the overall rate. The cuts
were chosen in order to roughly maximize
the statistical significances of different SEWS models
as given in Table~\ref{tableiv}.

The channel $W^\pm Z \rightarrow W^\pm Z$ is less interesting
than the $\wp\wm$, $\wp\wp$ and $ZZ$ channels 
for the following reasons. First, there is no direct $s$-channel
scalar contribution to the $W^\pm Z$ process, unlike the processes
$W^+W^-, ZZ$. Second, the LET-K $A(s,t,u)$ amplitude goes like $t/v^2$ which
gives a smaller cross section than that of 
$W^+W^- \rightarrow ZZ$ or $W^+W^+ \rightarrow W^+W^+$
where $A(s,t,u)\propto s/v^2$. 
Even if there is a $T_1$ Vector resonance,
the $\wp\wm$ final state will reveal a 
peak with the same strength as does the $\wpm Z$ final state,
with comparable backgrounds from the $T_0$ and $T_2$ channels
[compare Eqs.~(\ref{wwisospinii}) and (\ref{wwisospiniii})].
Finally, the cuts we have discussed
are not as successful for the $\wpm Z\to \wpm Z$
channel.  In particular, the $p_T(VV)$ and $\mu^\pm$ veto cuts are
no longer adequate to substantially suppress the (irreducible) background
associated the $\wpm\gam\to \wpm Z$ subprocess. 
Highly effective alternative cuts have not yet been identified. (See
Ref.~\cite{japanese}.) 

\begin{figure}[tbp]
\let\normalsize=\captsize   
\centering
\centerline{\psfig{file=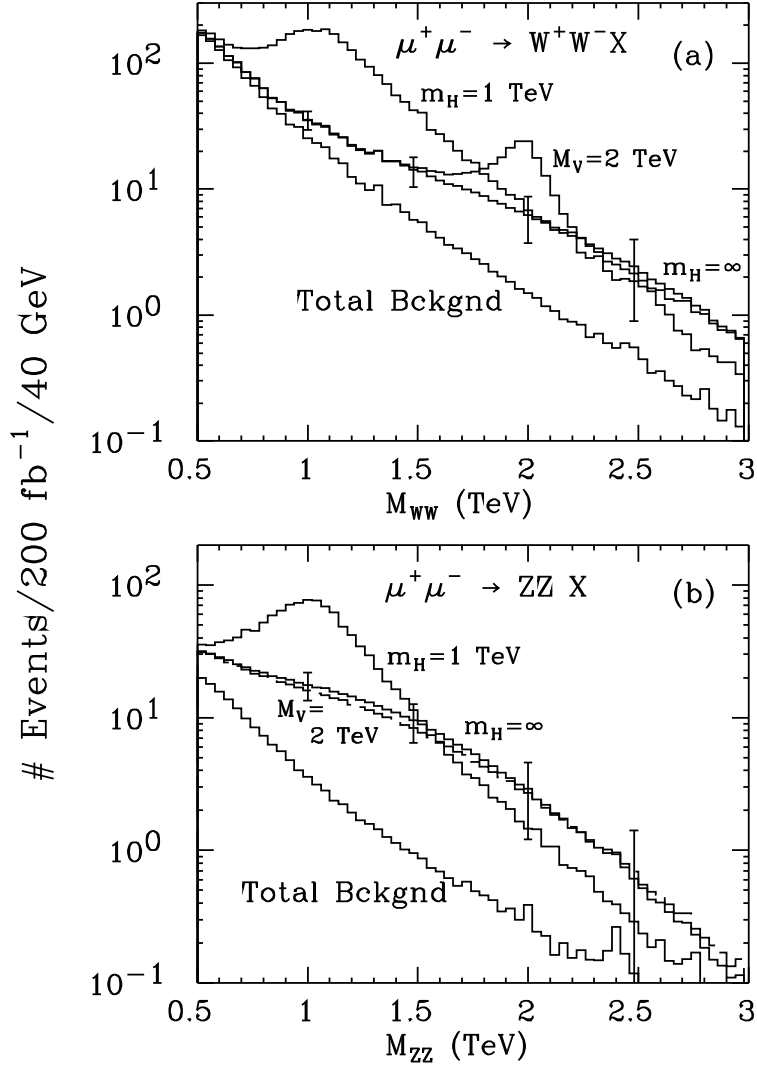,width=10cm}}
\begin{minipage}{12cm}       
\smallskip
\caption{{\baselineskip=0pt
Number of events at $\protect\rts=4\tev$ and $L=200\fbi$
versus $\mVV$ for SEWS models 
(including the combined backgrounds) and for the 
combined backgrounds alone in the 
(a) $\wp\wm$ and (b) $ZZ$ final states after imposing all cuts, I--III. 
Sample signals shown are: (i) the SM Higgs with $\mhsm=1\tev$;
(ii) the SM with $\mhsm=\infty$ unitarized via $K$-matrix techniques
(LET-K model); and (iii) the Vector model with $M_V=2\tev$
and $\Gamma_V=0.2\tev$. In the $ZZ$ final state the histogram for (iii)
falls just slightly lower than that for model (ii) at lower $\mVV$.
Sample statistical uncertainties for the illustrated 40 GeV bins are shown
in the case of the $\mhsm=\infty$ model.}}
\label{mvvfullcuts}
\end{minipage}
\end{figure}

\begin{figure}[tbp]
\let\normalsize=\captsize   
\centering
\centerline{\psfig{file=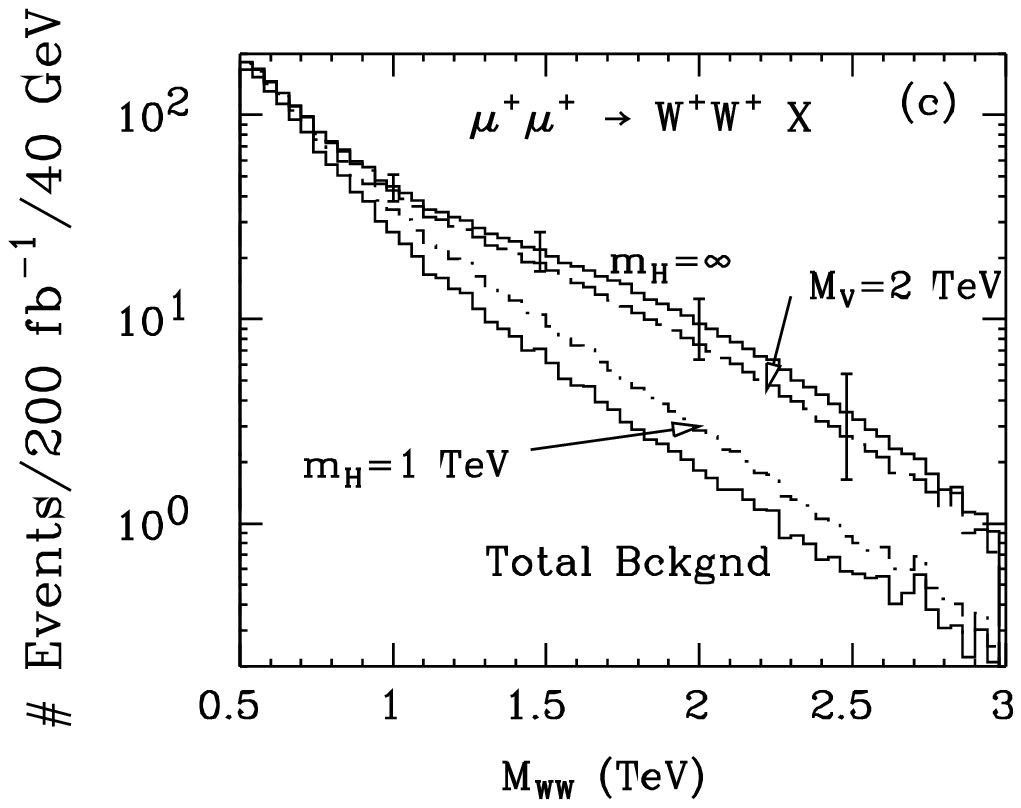,width=10cm}}
\begin{minipage}{12cm}       
\smallskip
\setcounter{figure}{4}
\caption{{\baselineskip=0pt (continued)
Events as a function of $\mVV$ for sample SEWS models 
(including the combined backgrounds) and for the 
combined backgrounds alone in the (c) $\wp\wp$
final state after imposing all cuts, I--III. See caption
for Fig.~\ref{mvvfullcuts}a-b.}}
\end{minipage}
\end{figure}

In Figs.~\ref{mvvfullcuts}a--c we compare
the $\mVV$ distributions in the $\wp\wm$, $ZZ$ and $\wp\wp$ final states
for various SEWS models 
(including the combined reducible and irreducible
backgrounds) to those for the combined background with
all cuts, I--III, imposed.
The SEWS models illustrated are the SM with $\mhsm=1\tev$, the unitarized
$\mhsm=\infty$ (LET-K) model, and a Vector model with $M_V=2\tev$
and $\Gamma_V=0.2\tev$. The numbers in Table~\ref{tableiv}
are obtained by integrating the distributions in these figures
over the specified $\mVV$ ranges,
where the signal event numbers are those obtained after
subtracting the background from the full SEWS model curves
(which include the combined background).
To indicate the accuracy with which the $\mVV$ distributions
could be measured,
the $L=200\fbi$, $\pm\sqrt N$ error bars associated
with several 40 GeV bins for the LET-K model are shown.

\begin{figure}[tbp]
\let\normalsize=\captsize   
\centering
\centerline{\psfig{file=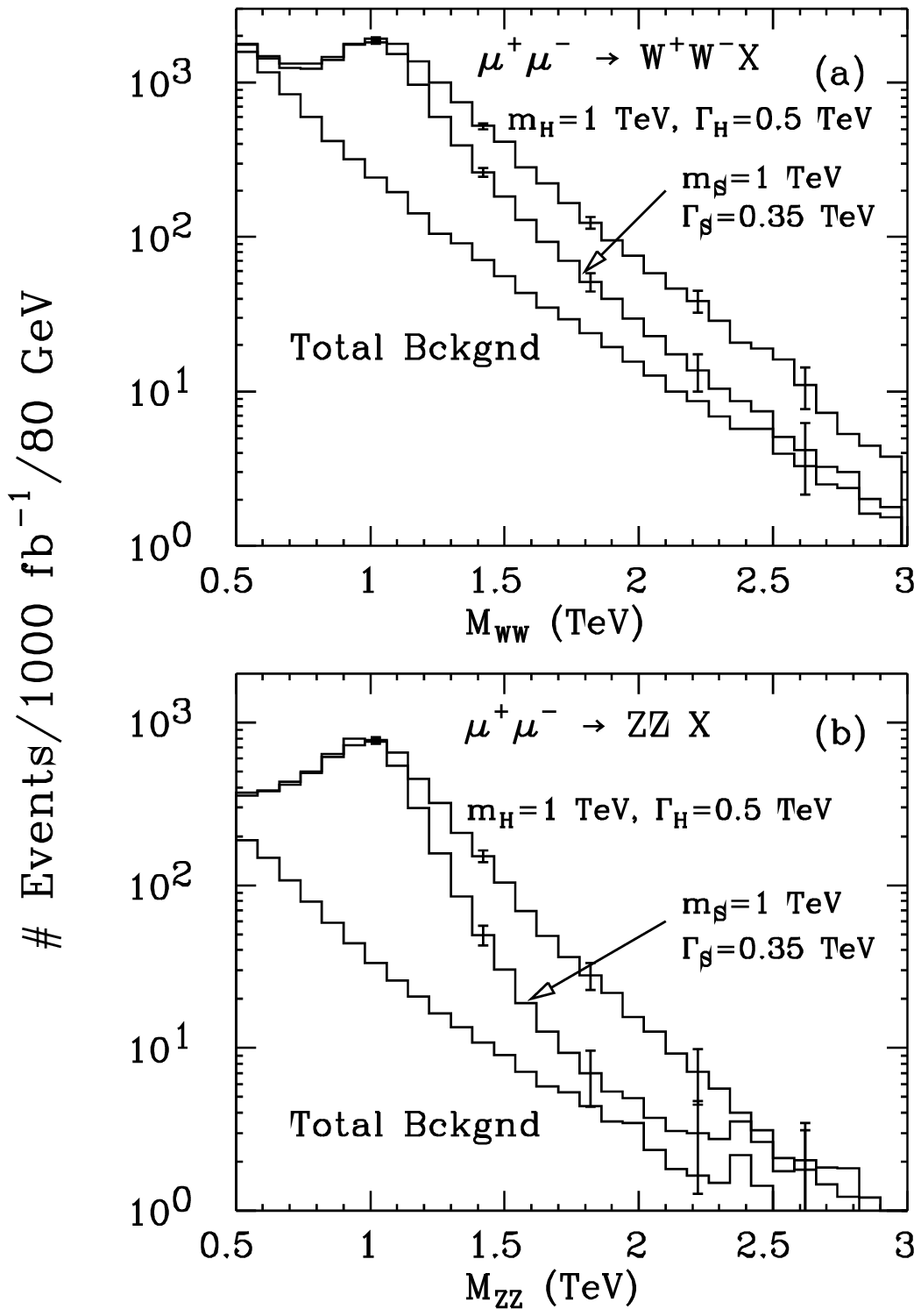,width=10cm}}
\begin{minipage}{12cm}       
\smallskip
\caption{{\baselineskip=0pt
Events versus $\mVV$ for two SEWS models 
(including the combined backgrounds) and for the 
combined backgrounds alone in the 
(a) $\wp\wm$ and (b) $ZZ$ final states after imposing all cuts, I--III. 
Signals shown are: (i) the SM Higgs with $\mhsm=1\tev$, $\Gamma_H=0.5\tev$;
(ii) the Scalar model with $M_S=1\tev$, $\Gamma_S=0.35\tev$.
Results are for $L=1000\fbi$ and $\protect\rts=4\tev$.
Sample error bars  are shown
at $\mVV=1.02$, $1.42$, $1.82$, $2.22$ and $2.62\tev$
for the illustrated 80 GeV bins.}}
\label{mvvfullcutswidth}
\end{minipage}
\end{figure}

From these plots and the sample error bars, 
it is apparent that, for any of the SEWS models investigated,
the expected signal plus background could be readily distinguished
from pure background alone on a bin by bin basis at better than $1\sigma$
all the way out to $\mVV=2.5\tev$ ($2\tev$) in the $\wp\wm$ and $\wp\wp$
($ZZ$) channels. Further,
the small $2\tev$ Vector model peak would 
be readily observed in the $\wp\wm$ channel and its absence
in the $ZZ$ and $\wp\wp$ channels would be clear.
Indeed, it would be feasible to determine
the width of either a scalar or a vector resonance with moderate accuracy.

Currently discussed designs for the $4\tev$ muon collider would
provide luminosity of $L=1000\fbi$ per year.
Even if this goal is not reached, one might reasonably anticipate
accumulating this much luminosity over a period of several years.
For $L=1000\fbi$, the accuracy with which the $\mVV$ distributions
can be measured becomes very remarkable.  To illustrate,
we plot in Figs.~\ref{mvvfullcutswidth}a-b, the signal plus background
in the $\mhsm=1\tev$, $\Gamma_H=0.5\tev$ SM and the $M_S=1\tev$,
$\Gamma_S=0.35\tev$ Scalar resonance model, and the combined background,
taking $L=1000\fbi$ and using an $80\gev$ bin size (so as to increase
statistics on a bin by bin basis compared to the $40\gev$ bin size
used in the previous figures). The error bars are almost invisible
for $\mVV\lsim 1.5\tev$, and statistics is more than adequate to
distinguish between the $\Gamma_H=500\gev$ SM resonance
and a $\Gamma_S=350\gev$ Scalar model at a resonance
mass of $1\tev$. Indeed, we estimate
that the width could be measured to better than $\pm 30\gev$.
Further, for such small errors we estimate that a vector resonance could
be seen out to nearly $M_V\sim 3\tev$.
This ability to measure the $\mVV$ distributions with high precision 
would allow detailed insight into
the dynamics of the strongly interacting electroweak sector.
Thus, if some signals for a strongly interacting
sector emerge at the LHC, a $\rts = 3-4\tev$ $\mm$ (or $\ee$, if feasible)
collider will be essential.

It is important to measure the $\mVV$
spectrum in all three ($\wp\wm$, $ZZ$ and $\wp\wp$) channels in order to
fully reveal the isospin composition of the model.
For instance, the Vector model and the LET-K model
yield very similar signals in the $ZZ$ and $\wp\wp$ channels,
and would be difficult to separate without 
the $\wp\wm$ channel resonance peak. More generally,
the ratio of resonance peaks in
the $ZZ$ and $\wp\wm$ channels would be needed to 
ascertain the exact mixture of Vector (weak isospin 1) and Scalar 
(isospin 0) resonances should they be degenerate.
Determination of the isospin composition
of a non-resonant model, such as the LET-K model, 
requires data from all three channels.
The $ZZ$ channel can only be separated from the $\wp\wm$ channel
if the jet energy resolution is reasonably good.

\section{SEWS Study using the Projection Procedure}

\begin{figure}[tbp]
\let\normalsize=\captsize   
\centering
\centerline{\psfig{file=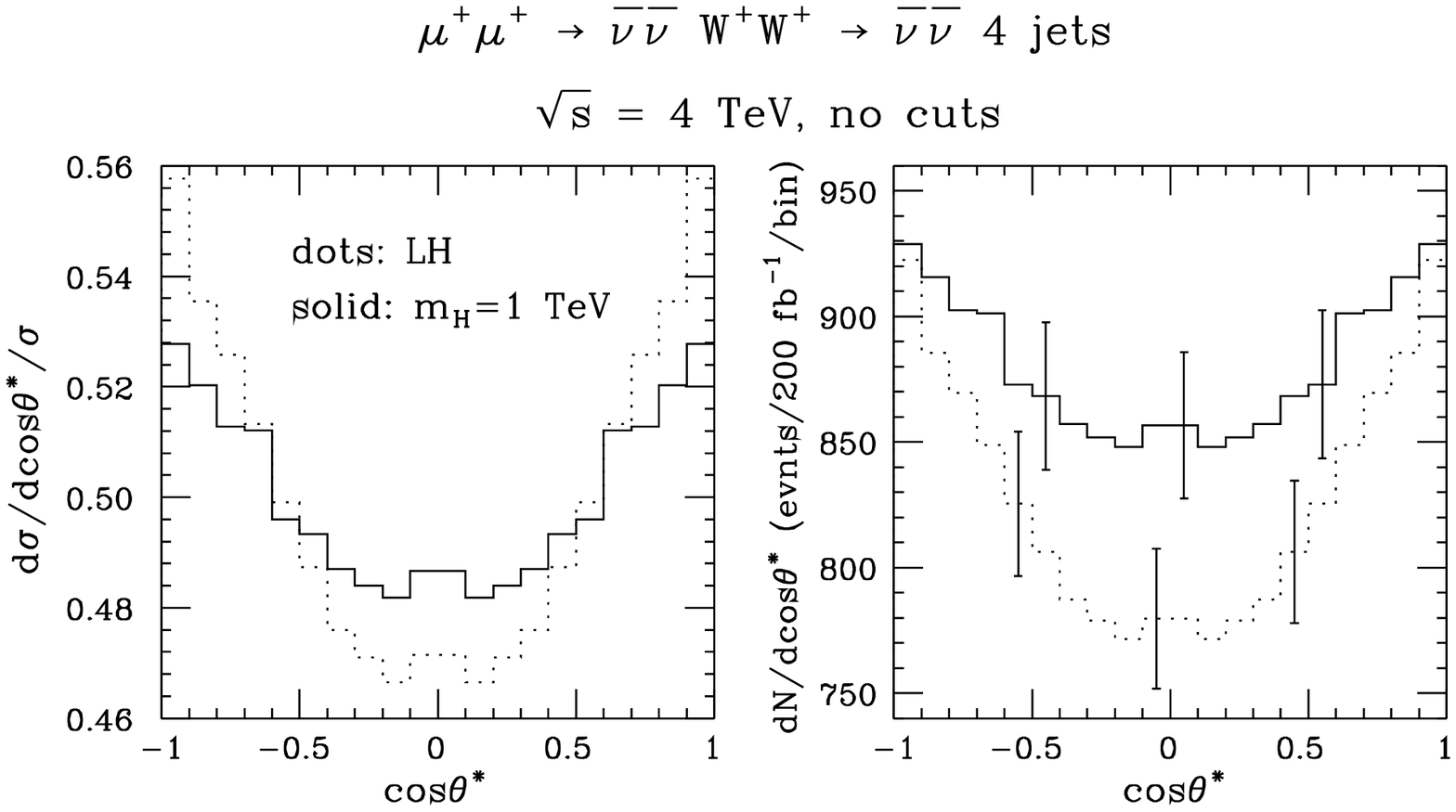,width=10cm}}
\begin{minipage}{12cm}       
\smallskip
\caption{{\baselineskip=0pt
Plots of normalized cross section shapes and $dN/d\cstar$ (for $L=200\fbi$)
as a function of the $\cstar$ of the $\wp$ decays in
the $\wp\wp$ final state.  Error bars for a typical $dN/d\cstar$
bin are displayed. For these two plots no cuts of any kind are performed.}}
\label{cosjnocut}
\end{minipage}
\end{figure}

\begin{figure}[tbp]
\let\normalsize=\captsize   
\centering
\centerline{\psfig{file=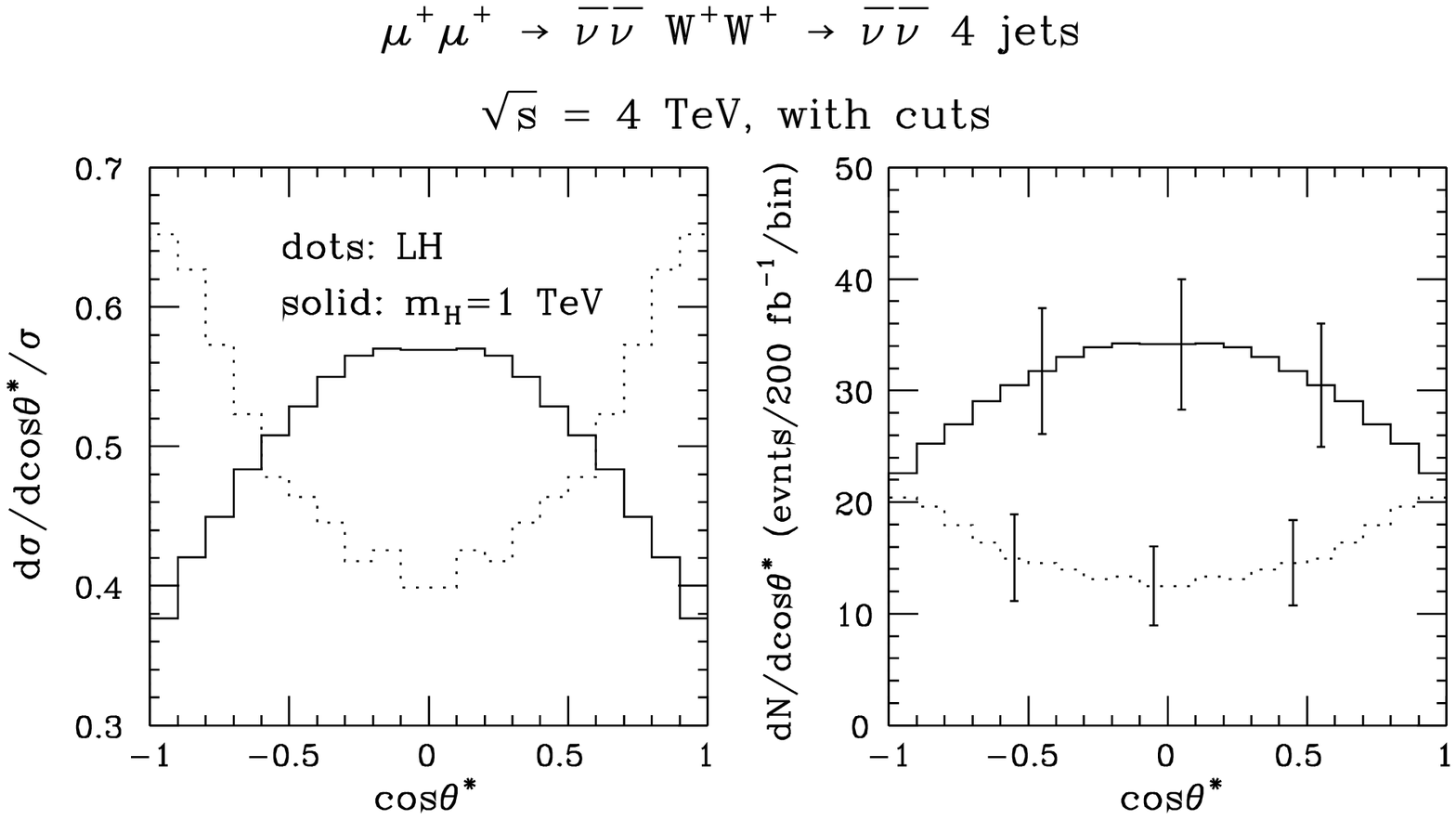,width=10cm}}
\begin{minipage}{12cm}       
\smallskip
\caption{{\baselineskip=0pt
Plots of normalized cross section shapes and $dN/d\cstar$ (for $L=200\fbi$)
as a function of the $\cstar$ of the $\wp$ decays in
the $\wp\wp$ final state.  Error bars for a typical $dN/d\cstar$
bin are displayed. For these two plots we require $\mVV\geq 500\gev$,
$p_T^V\geq 150\gev$, $|\cos\theta_W^{\rm lab}|<0.8$ and
$30\leq p_T^{VV}\leq 300\gev$.}}
\label{cosjcut}
\end{minipage}
\end{figure}

\indent\indent
It is advantageous to observe the $VV$ final state in the four-jet
mode in order to separate the $V_LV_L$,
$V_TV_L$ and $V_TV_T$ final states by angular
projection techniques. The angular distributions of interest
are those in $\cstari$ and
$\cstarii$, the cosines of the quark angles
in the $V_{1,2}$ rest frames, that we define relative to
the boost direction in the $V_1V_2$ center of mass. Since it is not
possible to distinguish quark from antiquark jets in the detector,
the configurations 
\begin{equation}
[\cstari,\cstarii]\,,\qquad [-\cstari,\cstarii]\,,\qquad
[\cstari,-\cstarii]\,,\qquad [-\cstari,-\cstarii]\,,
\label{fourconfig}
\end{equation}
must be averaged over. This automatically
avoids the problem of the ambiguous sign of $\cstari$ and $\cstarii$ 
in the rest frames of the two $V$'s.
Further, in the $\wp\wp$ and $ZZ$ modes
$V_1$ and $V_2$ cannot be distinguished; this applies
also to the $\wp\wm$ mode in the four-jet final state. 
Thus, our definition of 1 and 2 is arbitrary,
and the above four configurations much be averaged with their
$\cstari\leftrightarrow\cstarii$ counterparts.

Perhaps typical of the most difficult scenarios
would be the $\mup\mup\to\wp\wp \anti\nu\anti\nu$ channel 
in the $\mhsm=1\tev$ model.
As seen in Fig.~\ref{mvvfullcuts}c and discussed
with regard to Table~\ref{polarizationtable} the enhancement from strong
scattering is quite modest in this case. 
In this section, we focus on the ability of a projection analysis
to discriminate between this SEWS model and the \lh\ background.

Before proceeding with the projection analysis, it is useful
to simply examine some typical angular distributions. We construct
a one-dimensional `average' $\cstar$ plot by computing
$\cstari$ and $\cstarii$ for each event and entering
that event at $\cstar=\cstari$, $-\cstari$, $\cstarii$ and $-\cstarii$,
and dividing by four.
In Fig.~\ref{cosjnocut} the normalized
shapes ${1\over \sigma}{ d\sigma\over d\cstar}$ 
and the actual event number distributions
$dN/d\cstar$ (for $L=200\fbi$) for the \lh\ model and for $\mhsm=1\tev$
are compared before applying any cuts.
These same distributions are repeated in Fig.~\ref{cosjcut} after
imposing cuts I and II.
Even without any cuts, a distinct difference between the $\cstar$
distributions for the \lh\ model and $\mhsm=1\tev$ is observed.
The error bars displayed for typical $dN/d\cstar$ bins
make it clear that it would be easy to distinguish the two models
from one another using a combination of shape and normalization.
After the cuts, designed to enhance the $LL$ component of the cross
section, the shape distributions for the \lh\ model and for $\mhsm=1\tev$
are dramatically different. The error bars shown
in the $dN/d\cstar$ plot indicate
a discrimination between the two models of very high statistical
significance.  This will be quantified shortly.

In general, the amplitude as a function of the decay angles of the jets
from the two $V$'s is expressed in terms of a $VV$ helicity amplitude matrix
multiplied by appropriate helicity-dependent
$V$ decay amplitudes for the jets, summed over helicities.  
If the azimuthal angles of the jets in the rest frames of the two $V$'s
are integrated over, then the amplitude squared diagonalizes
yielding an expression of the form:
\begin{equation}
{d\sigma\over d\cstari\, d\cstarii}\equiv
\Sigma(\cstari,\cstarii)=\sum_{ij} \rho_{ij}f_i(\cstari)f_j(\cstarii)\,,
\label{siggeneral}
\end{equation}
where we have suppressed all kinematical variables except
$\cstari$ and $\cstarii$.
In Eq.~(\ref{siggeneral}), the $i,j$ indices are summed over $+$, $-$,
and $L$, and $f_+(z)\propto(1+z)^2$, $f_-(z)\propto(1-z)^2$, and 
$f_L(z)\propto(1-z^2)$.

\begin{figure}[tbp]
\let\normalsize=\captsize   
\centering
\centerline{\psfig{file=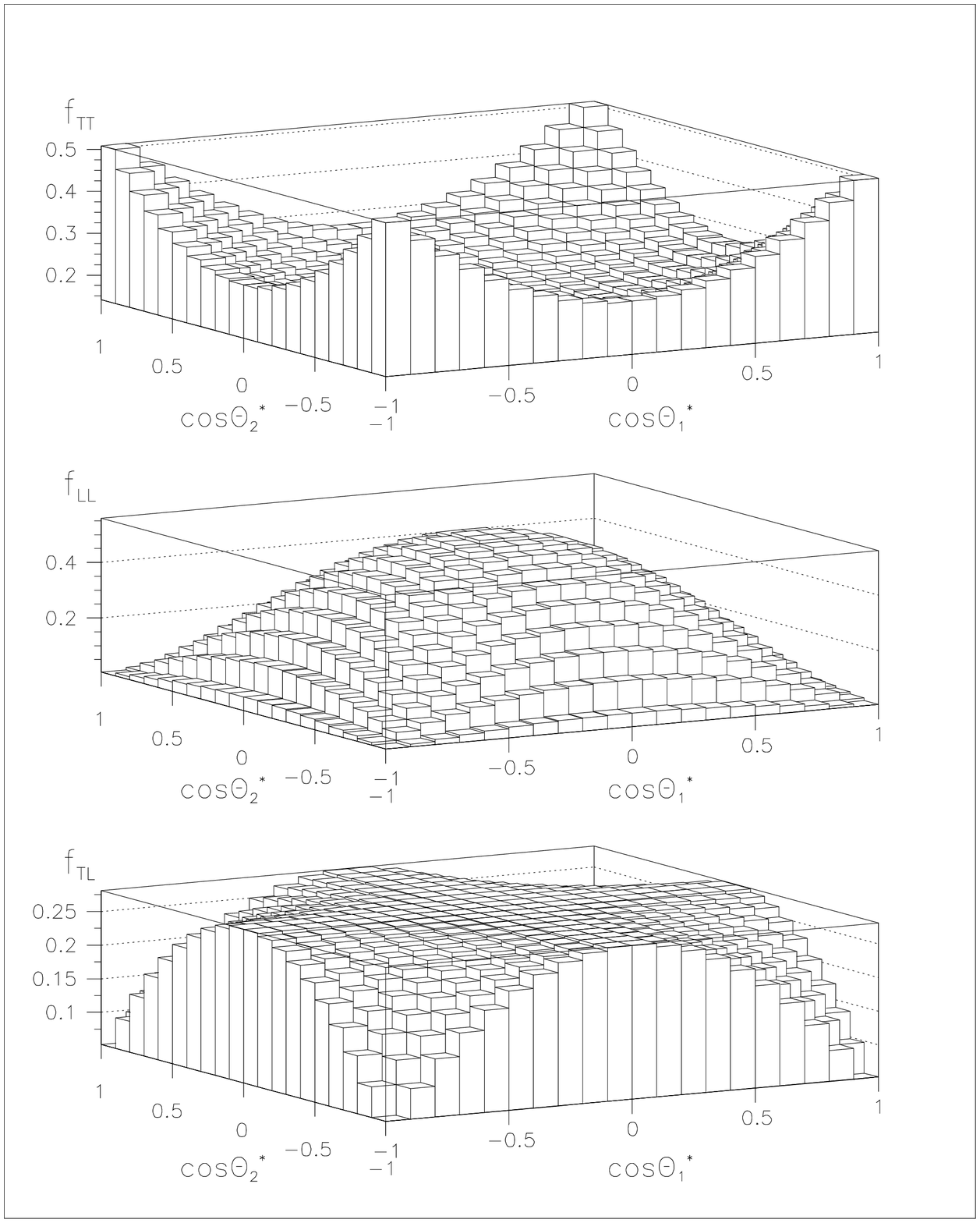,width=10cm}}
\begin{minipage}{12cm}       
\smallskip
\caption{{\baselineskip=0pt    The functions
$f_{TT}(\cstari,\cstarii)$, $f_{LL}(\cstari,\cstarii)$ and 
$f_{TL}(\cstari,\cstarii)$
[see Eq.~(\ref{fdefs})] are plotted as a function of $\cstari$ and
$\cstarii$ in two-dimensional parameter space.}}
\label{ttlltlfunctions}
\end{minipage}
\end{figure}

Because of our inability to distinguish quarks from and antiquarks,
we must bin in $\cstari$ and $\cstarii$ by entering
the weight for each event in the four bins specified in Eq.~(\ref{fourconfig}).
This results in a simplification due to the fact that
$f_+(z)+f_+(-z)=f_-(z)+f_-(-z)\propto 1+z^2$; $f_L(z)$ is not altered.
After symmetrizing over $\cstari\leftrightarrow\cstarii$
(due to our inability to distinguish $V_1$ from $V_2$),
the final form for the cross section as a function
of $\cstari$ and $\cstarii$ in the $ZZ$, $\wp\wm$ or $\wp\wp$ channels is
\begin{eqnarray}
\Sigma(\cstari,\cstarii) &=& \siga_{TT}f_{TT}(\cstari,\cstarii) + \siga_{LL}
f_{LL}(\cstari,\cstarii) \nonumber\\
\ &+& \siga_{TL}f_{TL}(\cstari,\cstarii)\,,
\end{eqnarray}
where 
\begin{eqnarray}
f_{TT}&=&f_T(\cstari)f_T(\cstarii)\,,\quad
f_{LL}=f_L(\cstari)f_L(\cstarii)\,,\nonumber\\
f_{TL}&=&{1\over 2}\left[f_T(\cstari)f_L(\cstarii)
+f_L(\cstari)f_T(\cstarii)\right]\,,
\label{fdefs}
\end{eqnarray}
with 
\begin{equation}
f_L(z)\equiv {3\over 4} (1-z^2)\,,\quad f_T(z)={3\over 8}(1+z^2)\,;
\end{equation}
the normalizations are chosen so that $\int_{-1}^{+1} f_i(z)\,dz=1$.
With this normalization, the $\siga_{LL,TT,TL}$ are the cross sections 
for $LL,TT,TL$ final states integrated over all
of $VV$ phase space (subject to cuts).
Thus, after integrating $\Sigma(\cstari,\cstarii)$ over 
$\cstari$ and $\cstarii$, the total
cross section is given by
\begin{equation}
\sigma_{\rm tot}=\sum_{i=TT,TL,LL}\sigma_i\equiv 
\siga_{TT}+\siga_{TL}+\siga_{LL}\,.
\label{sigdecomp}
\end{equation}
For later reference, the three functions of Eq.~(\ref{fdefs}) are plotted
in the two-dimensional $\cstari,\cstarii$ 
parameter space in Fig.~\ref{ttlltlfunctions}.
The goal of the projection analysis is to determine
the coefficients of these three
distinct two-dimensional distributions within a set of data that contains
an unknown mixture of them.

The optimal projection procedure (see Ref.~\cite{gghcp}) is 
to compute the matrix
\begin{equation}
M_{ij}\equiv \int {f_i(\cstari,\cstarii)f_j(\cstari,\cstarii))\over 
\Sigma(\cstari,\cstarii) }d\cstari\,d\cstarii \,,
\quad i,j=TT,LL,TL \,,
\label{mijdef}
\end{equation}
and the integrals
\begin{equation}
I_i\equiv \int f_i(\cstari,\cstarii) d\cstari\,d\cstarii\,,
\end{equation}
using the known $f_i$ and the experimentally measured $\Sigma$,
where the integrals are taken over $\cstari,\cstarii$.
The $I_i$ are equal to 1 in our normalization.
The coefficients $\siga_i$ are then determined 
as:\footnote{Here, and in all subsequent equations, the notation
$X^{-1}_{ij}$, where $X$ is any matrix, refers to the $i,j$
component of the inverse matrix, $X^{-1}$.}
\begin{equation}
\siga_i=\sum_j M_{ij}^{-1} I_j=\sum_j M_{ij}^{-1}\,,\quad i,j=TT,LL,TL\,.
\label{aiform}
\end{equation}

The above formulae assume that cuts are performed only on
the $V$'s and not on their jet decay products.
If significant cuts are performed on the jets then the procedure
becomes more subtle since the $\cstari$ and $\cstarii$ dependence
no longer necessarily factors from dependence on the other kinematical
variables.  The generalization is a well-defined extension
of that discussed above \cite{gghcp}.
We expect that the experimentally required jet cuts
will not significantly alter the results we shall
obtain without jet cuts, provided that the jet cuts are mild.
The analysis does become model-dependent if there are strong
correlations between other kinematic variables (especially $\mVV$)
and the ranges of $\cstari$ and $\cstarii$ that are accepted.

The expected experimental statistical 
errors in the projection determinations of the $\siga_i$
for a given model are determined in terms of
the covariance matrix $\langle\Delta \siga_i\Delta \siga_j\rangle$ defined by
\begin{equation}
V_{ij}\equiv \langle\Delta \siga_i \Delta \siga_j\rangle
={M_{ij}^{-1}\sigma_{\rm tot}\over N}\,,
\end{equation}
where $\sigma_{\rm tot}$ is defined in Eq.~(\ref{sigdecomp}),
$N$ is the total number of events expected (after cuts),
and $M_{ij}$ is computed from Eq.~(\ref{mijdef}) using
the model prediction for $\Sigma(\cstari,\cstarii)$.

Let us suppose that the model predicts values 
$\siga_i^0$ for the three coefficients.  (Note that the coefficient
$\siga_{TT}^0$ for any SEWS model is that predicted in the light Higgs
SM.) In a real experiment, Eq.~(\ref{aiform}) would yield values
$\siga_i^\star$ that are close to the $\siga_i^0$, and we would then
want to draw confidence-level ellipsoids about the $\siga_i^\star$.
We can approximate this procedure by assuming a given input
model with corresponding predictions for the $\siga_i^0$ and then
compute the $\Delta\chisq$ that would be associated
with values of the $\siga_i$ that differ from the input $\siga_i^0$:
\begin{equation}
\Delta\chisq=\sum_{i,j} (\siga_i-\siga_i^0)(\siga_j-\siga_j^0)V^{-1}_{ij}\,, \quad {\rm with}
~~V^{-1}_{ij}={M_{ij} N\over \sigma_{\rm tot} }\,,
\label{chisqdef}
\end{equation}
where, as above, $i,j=TT,LL,TL$. The confidence-level, $CL(\Delta\chisq)$
at which a fixed $\Delta\chisq$ ellipsoid can be said to contain
the the true values of the $\siga_i$ is then given in terms of
the cumulative distribution function $F$ (see Ref.~\cite{PDG}, Eq.~(16.22)) by
\begin{equation}
1-CL(\Delta\chisq)=F(\Delta\chisq,n)\,,
\end{equation}
where $n$ is the number of parameters: $n=3$ 
if all three $\siga_i$ are being considered.
If we are primarily interested
in the $TT$ and $LL$ coefficients, as will be case in the model 
considered in detail below, the correct procedure is
to take the $i,j=TT,LL$ submatrix of the covariance matrix $V_{ij}=M^{-1}_{ij}$,
invert it and apply Eq.~(\ref{chisqdef}) in the  $TT,LL$ ($n=2$)
parameter subspace. The 68.3\% and 90\% confidence-level
ellipses in this two parameter subspace are then defined by
$\Delta\chisq=2.295$ and $\Delta\chisq=4.606$, respectively.
The usual $1\sigma$ or 68.3\% confidence-level error on any one parameter $\siga_i$ 
without regard to other parameters is obtained by the one-parameter
version of the above procedure, and corresponds to $\Delta\chisq=1$, yielding
\begin{equation}
\Delta \siga_i=[M^{-1}_{ii}\sigma_{\rm tot}/N]^{1/2}\,.
\label{deltaadef}
\end{equation}
Below, we discuss only $\Delta \siga_i$ as defined above, but when
the experiment is actually performed it will be highly desirable
to construct the CL ellipsoids.\footnote{We note that the errors obtained
in the projection formalism closely approximate those that would result using 
a $\chisq$ minimization procedure in the $\siga_i$ for a given
known form of $\Sigma(\cstari,\cstarii)$ as a function of the $\siga_i$.}
We now analyze how successful this procedure can be in practice.

\begin{figure}[tbp]
\let\normalsize=\captsize   
\centering
\centerline{\psfig{file=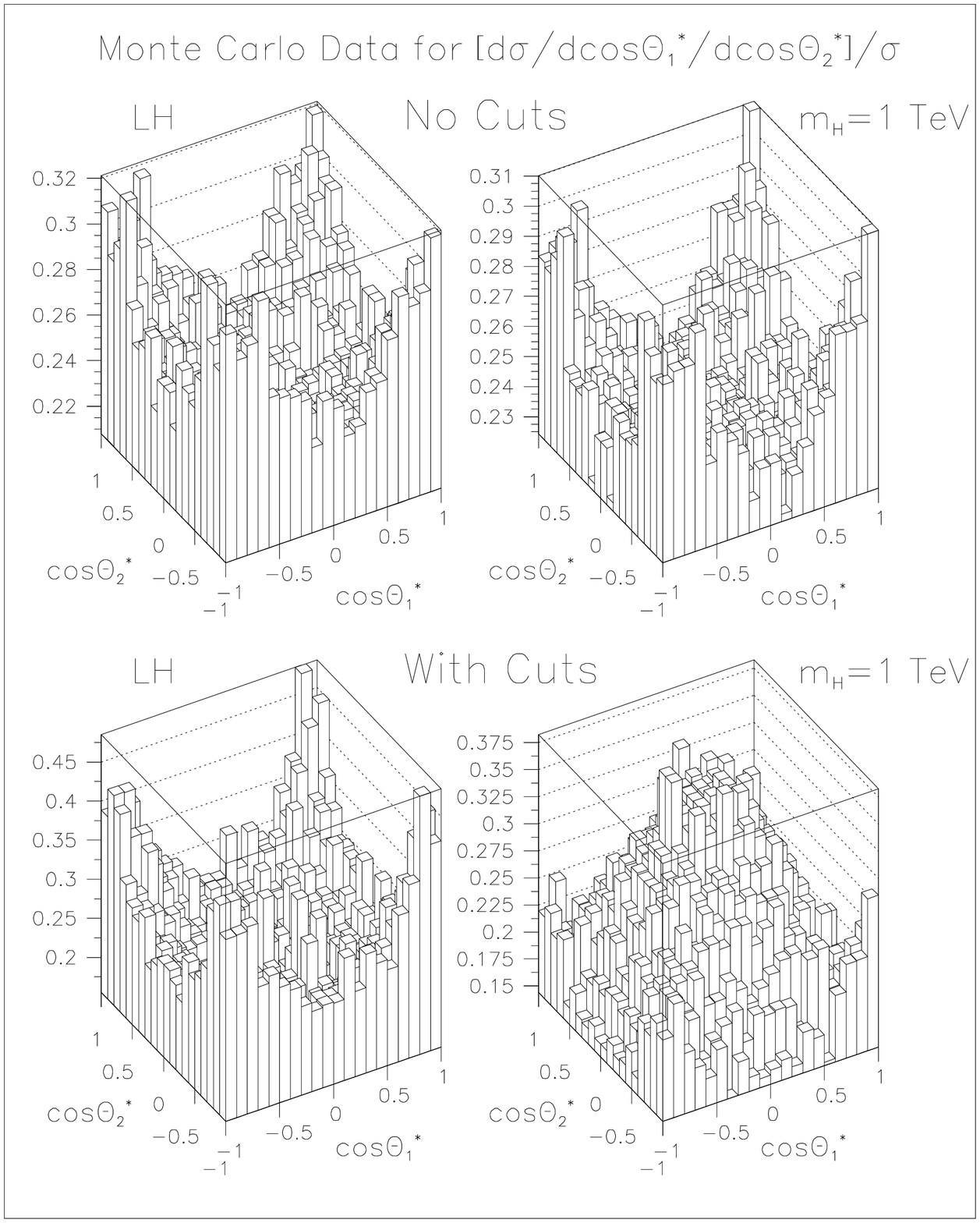,width=10cm}}
\begin{minipage}{12cm}       
\smallskip
\caption{{\baselineskip=0pt    Monte Carlo generated
prediction for ${1\over \sigma_{\rm tot}}{d\sigma\over d\cstari d\cstarii}$ 
in the $\wp\wp$ channel at $\protect\rts=4\tev$ 
in the two-dimensional $\cstari,\cstarii$ space for
four cases: (i) \lh, no cuts; (ii) $\mhsm=1\tev$, no cuts;
(iii) \lh, with cuts I and II; (iv) $\mhsm=1\tev$, with
cuts I and II.}} 
\label{2dmcdata}
\end{minipage}
\end{figure}

Let us turn to the full two-dimensional projection analysis.
The challenge is illustrated in Fig.~\ref{2dmcdata}.  There
we plot the Monte Carlo prediction for
${1\over \sigma_{\rm tot}}{d\sigma\over d\cstari d\cstarii}$ 
as a function of $\cstari$ and $\cstarii$ for the \lh\ model and for 
$\mhsm=1\tev$,
both before any cuts and after cuts I and II on the $\wp$'s.
Fluctuations in this figure are purely a result of the
Monte Carlo statistics, and are not meant to reflect actual
statistical errors in a typical experiment.
For the \lh\ model and no cuts, the 
Monte Carlo generated distribution mainly follows the expectations
for $f_{TT}$ (see Fig.~\ref{ttlltlfunctions}). For $\mhsm=1\tev$ and
no cuts, it is possible to observe some extra structure in
the vicinity of $\cstari\sim\cstarii\sim 0$ coming from a
$f_{LL}$ component.  The one-dimensional projection of 
these two-dimensional distributions, as plotted in Fig.~\ref{cosjnocut},
makes this difference clear.
After imposing cuts I and II, Fig.~\ref{2dmcdata} shows a dramatic
difference between the \lh\ and $\mhsm=1\tev$ models,
with the latter being heavily dominated by the $f_{LL}$ type
of structure -- the $f_{TT}$ component is mainly apparent in
the enhancements at the $|\cstari|=|\cstarii|=1$ corners.
Hence, our two-dimensional projection will
discriminate between the \lh\ and $\mhsm=1\tev$ models
at a very high level of statistical significance if
(after cuts are imposed) there are enough experimental events that 
the actual data resembles the Monte Carlo prediction.

\begin{table}[htb]
\centering
\caption[]{\label{projectiontable}\small
Percentage contributions of the $TT$, $TL$ and $LL$
$\wp\wp$ final states, for the \lh\ and $\mhsm=1\tev$ Standard Model cases
at $\rts=4\tev$, as computed theoretically and as obtained
from the projection analysis.}
\medskip
\begin{tabular}{|c|ccc|ccc|}
\hline
 \ & \multicolumn{3}{c|} {\lh} & \multicolumn{3}{c|}{$\mhsm=1\tev$}
\\
\ & $TT$ & $TL$ & $LL$ & $TT$ & $TL$ & $LL$ \\ \hline\hline
Theory/No Cuts & 59.2\% & 32.6\% & 8.2\% & 56.2\% & 30.1\% & 13.7\% \\ \hline
Projection/No Cuts & 59.4\% & 32.3\% & 8.3\% & 56.7\% & 29.1\% & 14.2\% \\ \hline
Theory/With Cuts & 85.9\% & 12.2\% & 1.9\% & 45.0\% & 5.0\% & 50.0\% \\ \hline
Projection/With Cuts & 87.0\% & 9.8\% & 3.2\% & 47.6\% & -1.3\% & 53.4\% 
\\ \hline
\end{tabular}
\end{table}

To test our projection analysis we have applied
the projection techniques outlined earlier to the (somewhat imperfect)
Monte Carlo generated distributions of
Fig.~\ref{2dmcdata} (before dividing by $\sigma_{\rm tot}$).
In Table~\ref{projectiontable}
we give the percentage contribution to the integrated cross section
deriving from $TT$, $TL$ and $LL$ final states both before
and after the cuts I and II, comparing results for the \lh\ case
to $\mhsm=1\tev$.  These percentages are simply computed
from the numbers in Table~\ref{polarizationtable}, which were
obtained by manually inserting the appropriate polarization
projectors into the Monte Carlo matrix elements.
Also presented in Table~\ref{projectiontable} are these same percentages
{\it as extracted from the Monte Carlo generated distributions
following the projection procedure outlined above}.
There is excellent agreement, except for the polarization combination
$TL$ when it is a very small fraction of the total cross section.
The success of the procedure is quite
remarkable given the substantial fluctuations in the Monte Carlo
distributions (Fig.~\ref{2dmcdata}) that we input. 
In particular, the projection procedure is successful in
demonstrating that the cross section increase in going from the \lh\ case
to $\mhsm=1\tev$ is primarily in the $LL$ mode,
{\it even in the case where no cuts are applied}.
It is important to note that the differences between the `Theory'
and `Projection' results in Table~\ref{projectiontable} are purely
those related to the limited accuracy of our Monte Carlo
integrations and have nothing to do with experimental errors.

\begin{table}[htb]
\centering
\caption[]{\label{projectionerrorstable}\small
Relative $1\sigma$ statistical errors, 
$\Delta \siga_i/|\siga_i|$ ($i=TT,TL,LL$),
expected for the projection technique in the $\wp\wp$
channel for the $\mhsm=1\tev$ SEWS model, assuming
$L=200\fbi$ at $\protect\rts=4\tev$.}
\medskip
\begin{tabular}{|c|ccc|ccc|}
\hline
 \ & \multicolumn{3}{c|} {\lh} & \multicolumn{3}{c|}{$\mhsm=1\tev$}
\\
\ & $TT$ & $TL$ & $LL$ & $TT$ & $TL$ & $LL$ \\ \hline\hline
Projection/No Cuts & 0.04 & 0.11 & 0.22 & 0.04 & 0.12 & 0.13 \\ \hline
Projection/With Cuts & 0.21 & 4 & 3 & 0.22 & 13 & 0.17 \\ \hline
\end{tabular}
\end{table}

In order to determine the relative ($1\sigma$)
statistical errors expected for the different cross section components
when employing the projection technique to real experimental data,
we calculate $\Delta \siga_i/|\siga_i|$ with $\Delta \siga_i$ 
using Eq.~(\ref{deltaadef}) with $N$ as predicted for $L=200\fbi$
and using the exactly computed $M_{ij}$ for the model in question.
These relative errors are presented in Table~\ref{projectionerrorstable}.
For those cross section components that are a substantial fraction
of the total, the relative statistical errors are quite good.
They would be a factor of 2.23 better for $L=1000\fbi$.
We give a few examples:
\begin{itemize}
\item
For $L=200\fbi$, the $2\sigma$ upper limit on $\sigma_{LL}$ [see
Eq.~(\ref{sigdecomp})] for no cuts (cuts) and the \lh\ model is $7.86\fb$ 
($0.24\fb$) while the
$2\sigma$ lower limit on $\sigma_{LL}$ for no cuts (cuts) and $\mhsm=1\tev$
is $8.76\fb$ ($1.0\fb$). 
\item
In the cuts case, for $L=200\fbi$, the $4\sigma$ upper limit
on $\sigma_{LL}$ for the \lh\ model is $0.48\fb$ and the $4\sigma$ lower
limit on $\sigma_{LL}$ for $\mhsm=1\tev$ is also $0.48\fb$. 
\end{itemize}
Thus, especially by applying cuts, a high level of statistical discrimination
between the $\mhsm=1\tev$ and the \lh\ models is possible.
We re-emphasize that this is one of the most difficult cases
that we could have considered.  Statistical discrimination for
most other models and channels would be very dramatic indeed.

In order to obtain a more detailed picture of
the SEWS model function $A(s,t,u)$
defined earlier, it would be very advantageous if 
the above analysis could be applied on a bin-by-bin
basis as a function of $\mVV$. For $\mhsm=1\tev$ in the $\wp\wp$ channel,
we estimate that the minimum bin size needed to retain adequate
statistics for $\mVV\lsim 2\tev$
in the projection analysis at $L=1000\fbi$ is $\sim 250\gev$. Further work
on this type of analysis is in progress. 

It is important to assess the impact of a realistic experimental
detector environment upon the projection procedure.
Since this is highly detector dependent, we only make some general
comments.
\begin{itemize}
\item Jet cuts should not greatly decrease the viability
of the projection procedure. If there is significant
distortion due to the non-factorization
of $\cstari$ and $\cstari$ dependencies from other kinematical variables,
then the generalized procedure
(discussed in general terms in \cite{gghcp}) must be followed,
and the extraction of the $\sigma_i$ would become somewhat model-dependent.
\item We have investigated the extent to which smearing of the
jet energies results in a deterioration of the procedure.
This can affect the experimental determination of
$\cstari$ and $\cstarii$ because the smearing
is in the laboratory frame and the smeared momenta
are then used to determine the boosts required to go to
the $\wp\wp$ rest frame and then to the individual $\wp$ rest frames
where the angles are ultimately defined.
For jet energy resolution of order $\Delta E/E\sim 50\%/\sqrt E \oplus 2\%$,
the effect is smaller than the Monte Carlo statistics that
we were able to achieve in our program.
\item An important experimental issue is the ability of the detector
to properly resolve the two jets coming from a given $\wp$.  
On the average, they are separated at $\rts=4\tev$ by about 
$17^\circ$ in the laboratory. The detector
must be designed with this in mind.  Failure to achieve good
separation of the two jets would mean that the projection
procedure could not be employed. Detectors being discussed
will have sufficient segmentation that good separation should be possible.
\item A closely related issue is 
the uncertainty in the experimental determination of the
angles of the jets in the laboratory frame. Errors
in these angles could possibly lead to
a distortion in the determination of $\cstari$ and $\cstarii$
that is larger than that from simple energy smearing.  
\end{itemize}
Because the last two items are so detector dependent, we have not
attempted a detailed study.

\section{Summary and Conclusion}

\indent\indent
Achieving $VV$ scattering subprocess energies above $1-2\tev$
is critical for studies of 
strongly interacting electroweak sector (SEWS) models, and is
only possible with high event rates at lepton-antilepton 
($\ee$ or $\mm$ colliders) or quark-antiquark (hadron collider)
subprocess energies of order $3-4\tev$. 
Consequently, a muon collider facility with center
of mass energy $\rts\sim 3-4\tev$ and luminosity
$L=200-1000\fbi$  allowing both $\mup\mum$ and $\mup\mup$
(or $\mum\mum$) collisions would be a remarkably powerful machine for
probing a strongly interacting electroweak sector (SEWS).  
The LHC or a lower energy $\epem$ collider would not be competitive.

Event rates for even the weakest of the model signals studied are such
that the $\mVV$ distributions could be quantitatively
delineated, thereby providing a direct measurement of the underlying
strong $VV$ interaction amplitude as a function of the $VV$ subprocess
energy and strong differentiation
among various possible models of the strongly interacting electroweak
sector. 

Statistics are even sufficient that a model-independent
projection analysis can be applied to $\mVV$ integrated data
(either before or after cuts) to isolate the $TT$, $TL$ and $LL$
components of the cross section. Employing
the projection techniques with acceptance cuts imposed
to extract the $LL$ component of the $\wp\wp$ cross section,
for $L=200\fbi$  we found that the $4\sigma$ statistical upper limit
for the light Higgs Standard Model prediction is slightly lower
than the $4\sigma$ lower limit for the $\mhsm=1\tev$ result. This
level of statistical discrimination between the models
is remarkable, considering that the $\mhsm=1\tev$ cross section in
the $\wp\wp$ channel
is one of the most difficult channels and models for isolating
the $LL$ cross section. For most SEWS models, the statistical
level at which a model-independent extraction of the $LL$
cross section could be demonstrated to be inconsistent with
the light Higgs Standard Model expectation would be much larger.
Comparison of $LL$ cross sections in the $\wp\wm$, $ZZ$ and $\wp\wp$
final states with one another and with model expectations
would single out a small class of viable SEWS theories.
We are optimistic that the $LL$ cross section
could be extracted for many SEWS models and channels 
even on a bin-by-bin basis in $\mVV$, 
with a model and channel dependent bin size of order $250\gev$ (in the range
$\mVV\lsim 2\tev$), if $L=1000\fbi$ of data is available.
This would further delineate the correct theory
underlying the strong electroweak interactions.

Thus, if evidence for a strongly interacting
electroweak sector emerges from LHC or NLC data, construction of a 
high luminosity, high
energy muon collider (or electron collider, if feasible) should
be given the highest priority.

\bigskip\goodbreak
\begin{center}
{\bf\uppercase{Acknowledgments}}
\end{center}

This work was supported in part by the U.S.~Department of Energy  
under Grants No.~DE-FG02-95ER40896, No.~DE-FG03-91ER40674 and  
No.~DE-FG02-91ER40661. 
Further support was provided
by the University of Wisconsin Research
Committee, with funds granted by the Wisconsin Alumni Research  
Foundation, and by the Davis Institute for High Energy Physics.

\begin{center}
{\bf\uppercase{Appendix}}
\end{center}

\addcontentsline{toc}{section}{Appendices}
\section*{$K$-matrix Unitarization for $WW$ scattering Amplitudes}
\renewcommand{\theequation}{A.\arabic{equation}} 
\setcounter{equation}{0}

\indent\indent
$WW$ scattering amplitudes in SEWS models violate unitarity at high
energies, especially for the non-resonant scattering, such as $ZZ$
and $W^+W^+$, final states.
One therefore must unitarize them in some specific scheme to obtain
physical results. For simplicity, we have taken the $K$-matrix
unitarization scheme \cite{hikasa}. Namely, for a given partial wave amplitude, 
$a_l$, we unitarize it by the following replacement:
\begin{equation}
\label{EQ:K}
a_l \rightarrow \frac{a_l}{1-ia_l}.
\end{equation}
The partial wave amplitude $a_l$ before the unitarization
is obtained from the isospin amplitudes $T_I$,
\begin{equation}
a_l^I = \frac{1}{64\pi }\int_{-1}^{1} d\cos\theta P_l(\cos\theta) T_I\,.
\end{equation}
In turn, $T_I$ is given by the fundamental amplitude function 
$A(s,t,u)$ as discussed in Sec.2.

Inversely,
\begin{equation}
T_I = 32 \pi \sum_{l=0}^{\infty} (2l+1) P_l(\cos\theta) a_l^I.
\label{EQ:INV}
\end{equation}
Applying Eq.~(\ref{EQ:K}) to $a^I_l$, the amplitudes $T_I$ and
the physical scattering amplitudes in 
Eqs.~(\ref{wwisospini})-(\ref{wwisospiniv}) are thus unitarized.

\subsection{Unitarization of LET amplitudes}

\indent\indent
The fundamental amplitude function $A(s,t,u)$ according to the 
low energy theorem is given by
\begin{equation}
A(s,t,u) = s/v^2.
\end{equation}
Before the unitarization, the $T_I$'s are
\begin{equation}
T_0 = \frac{2s}{v^2},  \quad
T_1  =  \frac{s}{v^2}\cos\theta,  \quad
T_2  = -\frac{s}{v^2},
\label{EQ:TLET}
\end{equation}
where $\cos\theta=(t-u)/s$. Correspondingly,
\begin{equation}
a_0^{I=0} = \frac{1}{16\pi} \frac{s}{v^2},  \quad
a_1^{I=1}  =  \frac{1}{96\pi}\frac{s}{v^2},  \quad
a_0^{I=2}  = -\frac{1}{32\pi} \frac{s}{v^2}.
\label{EQ:ALET}
\end{equation}
Inversely from Eq.~(\ref{EQ:INV}),
\begin{equation}
T_0 = 32\pi a^{I=0}_0,  \quad
T_1  =  96\pi a^{I=1}_1\cos\theta,  \quad
T_2  = 32\pi a^{I=2}_0.
\label{EQ:TINV}
\end{equation}
The amplitudes are thus unitarized by applying Eq.~(\ref{EQ:K})
to $a^I_l$.

\subsection{Unitarization for the Vector Model}

\indent\indent
We now present the unitarization procedure in the Vector Model
for the non-resonant channels
\begin{equation}
W^+W^- \rightarrow ZZ, \quad {\rm and} \ \  W^+W^+ \rightarrow W^+W^+.
\label{wwzz}
\end{equation}
The fundamental amplitude function $A(s,t,u)$ in the Vector Model
is given by \cite{bbcghlry}
\begin{equation}
A(s,t,u) = \frac{s}{4v^2} (4-3\alpha)
+ \frac{\alpha M_V^2}{4v^2} 
(\frac{u-s}{t-M^2_V} + \frac{t-s}{u-M^2_V}) .
\end{equation}
where $\alpha$ is a model parameter \cite{bbcghlry}.
We need only $T_0$ and $T_2$ to evaluate the scattering
amplitudes of Eq.~(\ref{wwzz}) and we find that
\begin{equation}
T_0 = 2 A(s,t,u)\ ,  \quad
T_2  = -A(s,t,u)\ .
\label{EQ:TVEC}
\end{equation}
If we assume $s$-wave dominance, then the $l=0$ partial
wave amplitudes can be expressed as
\begin{equation}
a_0^{I=0} = \frac{1}{16\pi} \frac{s}{v^2} F(\alpha,M^2_V,s),  \quad
a_0^{I=2}  = -\frac{1}{32\pi} \frac{s}{v^2} F(\alpha,M^2_V,s),
\label{EQ:AVEC}
\end{equation}
where 
\begin{equation}
F(\alpha,M^2_V,s) = 1-\frac{\alpha}{2}(\frac{3}{2} +
\frac{M^2_V}{s}) + \frac{\alpha}{2} (2+ \frac{M^2_V}{s})
\frac{M^2_V}{s}\ln(1+\frac{s}{M^2_V}).
\label{EQ:FUN}
\end{equation}
With the relations in Eq.~(\ref{EQ:TINV}),
the amplitudes can be unitarized by applying Eq.~(\ref{EQ:K})
to $a^I_l$ in Eq.~(\ref{EQ:AVEC}).

\end{document}